\documentclass[modern]{aastex701}
\usepackage{amsmath}

\newcommand{\Figref}[1]{Figure \ref{#1}}
\newcommand{\figref}[1]{Figure \ref{#1}}
\newcommand{\Secref}[1]{Section \ref{#1}}
\newcommand{\secref}[1]{Section \ref{#1}}
\def\d{\mathrm{d}}
\newcommand{\D}{\mathrm{D}}
\newcommand{\pp}[3][]{{\partial^{#1} #2 \over \partial #3^{#1}}}

\newcommand{\ubb}{\bm{u}}

\usepackage[scaled=1.0, varg]{newtx}
\usepackage{bm}

\begin{document}

\title{Mechanisms of Superrotation in Slowly-Rotating and Tidally-Locked Planets}

\author[orcid=0000-0002-4116-973X,sname='Nicolas']{Quentin Nicolas}
\affiliation{Institute for Atmospheric and Climate Science, ETH Zurich, Zurich, Switzerland}
\email[show]{quentin.nicolas@env.ethz.ch}  

\author[orcid=0000-0002-5971-8995, sname='Vallis']{Geoffrey K. Vallis} 
\affiliation{Department of Mathematics and Statistics, University of Exeter, Exeter, UK}
\email{gkvallis@e\gmail.com}

\begin{abstract}
Superrotation is a common feature of quickly rotating gas giants (e.g., Jupiter), slowly rotating planetary bodies (e.g., Titan), and tidally-locked planets. In this paper we compare and contrast the mechanisms of superrotation in slow rotators and tidally-locked planets. We cover a wide range of planetary properties, varying in particular the thermal Rossby number $Ro_T$ (controlled by planetary size, rotation rate, and instellation) and a radiative relaxation timescale $T_\mathrm{rad}$ (which parameterizes atmospheric optical thickness).
We use a two-level model that contains the principal mechanisms for superrotation in both regimes yet remains analytically tractable. Linearizations of the model elucidate the behavior of superrotation-inducing eddies. In tidally-locked planets a Matsuno--Gill-like structure organizes the eddy effects but of itself is insufficient to produce superrotation; baroclinicity and low-level drag are additional essential ingredients. Nonlinear integrations further explore the superrotating regimes and exhibit significant time variability even in statistical equilibrium. Not all tidally-locked regimes superrotate: subrotation arises at high $T_\mathrm{rad}$ (optically thick atmospheres) and weak low-level drag. 
On axisymmetrically-forced slow rotators, superrotation is ubiquitously linked to a previously identified Rossby--Kelvin instability. Perhaps surprisingly, the instability itself is also linked to the spinup of superrotation in some tidally-locked regimes. Finally, we explore the continuous transition in the mechanisms of superrotation from axisymmetrically-forced to tidally-locked planets by applying a progressively stronger asymmetric equatorial forcing. The Matsuno--Gill pattern quickly dominates over traveling planetary Rossby--Kelvin waves in forcing superrotation, although both mechanisms can coexist. These results provide a unified view of superrotation mechanisms across a wide range of planetary bodies.
\end{abstract}




\section{Introduction} 
Axisymmetric motion in planetary atmospheres cannot produce an angular momentum maximum away from a surface or an interior quiescent layer, a consequence of Hide's result \citep{Hide1969, Vallis2017}. Hence, equatorial winds must be zero or retrograde when the flow is purely axisymmetric. Despite this constraint, Venus, Jupiter, Saturn, Titan, and many planetary atmospheres beyond the solar system have prograde equatorial winds; they \textit{superrotate}. On these planets, non-axisymmetric wave processes presumably flux momentum from extratropical and subtropical regions towards the equator.

At least three broad classes of planets exhibit superrotation: fast-rotating gas giants (e.g., Jupiter and Saturn), slowly rotating terrestrial planets (e.g., Venus and Titan), and some tidally-locked planets. Most observed exoplanets are tidally-locked, if only because current detection techniques favor planets orbiting close to their host star where tidal stresses are expected to rapidly bring to a state of tidal locking  \citep{Barnes2017}. This last group potentially contains both terrestrial planets and gas giants (the latter often referred to as hot Jupiters), and possibly other classes of planets.

Terrestrial planets are commonly `shallow', meaning that the depth of the layer in which atmospheric flows take place is small compared to the planet's radius. For such atmospheres, the primitive equations and shallow-water equations are very useful tools. Shallow flow can also occur in gas giants if the flows are confined to an upper stratified layer. On cold giants, such as Jupiter and Saturn, the flows seem to originate in (or at least extend into) a deeper convective layer \citep{Kaspi2020} and there the flow aligns with the rotation axis and shallow atmospheric models are inappropriate. On hot Jupiters, however, the internal heat flow is likely to be very weak compared to the stellar irradiation. The upper atmosphere may then be stably stratified and the flow shallow (as, for example, assumed by \cite{Showman2008}). One might crudely estimate this depth as a density scale height, which is typically a factor of 20 or more smaller than the planetary radius. If these assumptions are applicable (even if not exact), both slowly rotating terrestrial planets and tidally-locked hot Jupiters can be modeled in a single framework, that describing shallow atmospheres -- or at least that is the approach we shall take here. (We of course do not discount the possibility that hot Jupiters have interesting internal dynamics.)

Given this, the question arises as to whether tidally-locked and slowly-rotating but axi-symmetrically forced planets (henceforth referred to just as `slow rotators') share a common mechanism producing superotation, or whether the mechanisms are fundamentally different. Various classes of waves have been proposed to drive superrotation in both cases. For the slow-rotators an instability of the zonally symmetric basic state is likely required to create eddies, and one such (the Rossby--Kelvin -- or RK -- instability) arises when midlatitude Rossby waves phase lock with equatorial Kelvin waves and produce momentum-converging wind patterns \citep{Iga2005, Wang2014}. For tidally-locked planets, the steady linear response to non-axisymmetric heating, namely a Matsuno--Gill-like pattern, can itself spin-up superrotation if vertical momentum transport (associated with a vertical mass transport due to heating) from a lower quiescent layer is also taken into account \citep{Showman2011, Tsai2014}.

While shallow water models have been immensely valuable in the exploration of mechanism, producing realistic superrotation with these models has proved challenging. The vertical momentum transport parameterization employed by \cite{Showman2011} requires some amount of retrograde flow at the equator for superrotation to exist.
The parameterization prevents the emergence of pan-equatorial superrotating flows such as found by global circulation models (GCMs) of tidally-locked planets \citep[e.g.,][]{Showman2009, Pierrehumbert2019, Lewis_etal21}. For slow rotators, \cite{ZuritaGotor2018} have shown that a 1.5-layer shallow water model, even with vertical momentum transport, struggles to produce realistic superrotation despite representing the RK instability. 

The present work exploits a modeling framework that contains the simplest physical processes needed to produce more realistic superrotation (i.e., characteristic of that observed or simulated by full GCMs) in slow rotators and tidally-locked planets. We seek to answer the following questions: Is it likely that all tidally-locked planets superrotate? More specifically, for what sets of parameters (size, rotation rate, insolation, surface drag, etc.) does superrotation occur? Are the mechanisms of superrotation on tidally-locked planets and slow rotators related? Can both co-exist, or does one naturally dominate, and if so under what circumstances?

\Secref{sec:methods} presents the two-level model used to address these questions. \secref{sec:theory} explores the theoretical underpinnings of superrotation-inducing eddies on tidally-locked planets and slow rotators in that model. In \secref{sec:nonlinear}, the fully nonlinear version of the model is integrated to a statistically steady state for a wide range of planetary parameters, for both tidally-locked planets and slow rotators, to map the appearance of superrotation and the speed of the equatorial jet on two key parameters: a thermal Rossby number, and a nondimensional thermal relaxation scale.  \Secref{sec:transition} presents a continuum of simulations that transitions from tidally-locked to axisymmetrically-forced states. These are used to probe the interplay between the momentum-converging eddies characteristic of both states. Concluding remarks can be found in \secref{sec:conclusions}. 

\section{Methods} \label{sec:methods}
Superrotation cannot be sustained in the absence of vertical momentum transport if the circulation is symmetric about the equator at all times \citep[e.g.,][]{Showman2010}. Arguably, the simplest models of superrotation are ``1.5-layer'' shallow water models, where vertical momentum transport from a lower quiescent layer is parameterized. There are two drawbacks to such models: they do not contain baroclinic instabilities (a crucial process in generating Rossby waves that can cause westward equatorial acceleration, impeding superrotation), and they have  difficulties in producing superrotation in slowly rotating planets \citep[e.g.,][]{ZuritaGotor2018}. 

\begin{figure}[!t]
\begin{center}
 \plotone{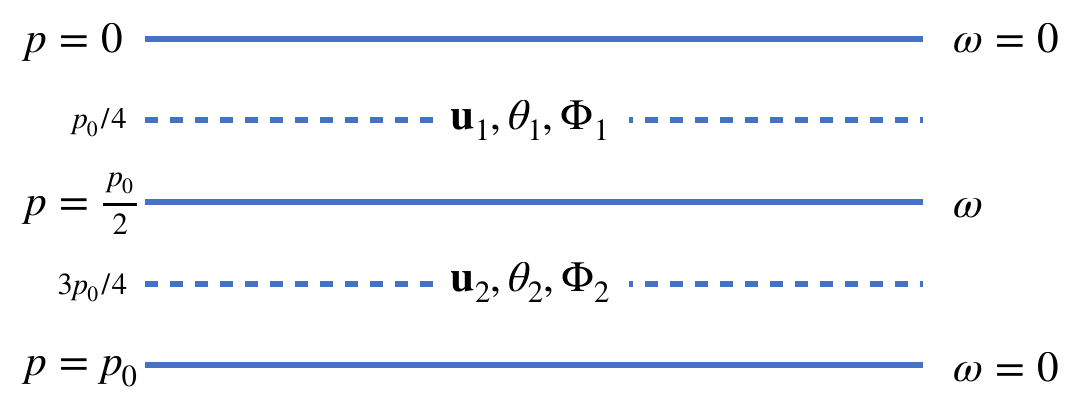}
 \caption{Vertical discretization of the 2-level atmospheric model. A staggered grid is employed, where pressure velocity is defined on the full levels $p=0, p_0/2, p_0$, and horizontal velocities, geopotential heights and potential temperatures are defined on the half levels $p_1 = p_0/4$ and $p_2 = 3p_0/4$. \label{fig:2levsketch}}
\end{center}
\end{figure}
We choose here to use a rather different model, based on a vertical discretization of the hydrostatic primitive equations: the two-level model used by \cite{Held1978}, also used in the studies of superrotation by \cite{Suarez1992} and \cite{Saravanan1993}. The vertical discretization is illustrated in \figref{fig:2levsketch}. The lowest pressure level of the model, $p_0$, is taken as the mean surface pressure on a terrestrial planet. On a gas giant, it is the pressure level at which horizontal flows become negligible,compared to weather-layer flows. The governing equations for momentum, potential temperature, hydrostasy, and continuity read:
\begin{align}
\pp {\ubb_i} t + \bm{u}_i\cdot\nabla\bm{u}_i + \omega(\bm{u}_2 - \bm{u}_1) + f \bm{k}\times\bm{u}_i &= - \nabla \Phi_i -  \delta_{i2}\dfrac{\bm{u}_i}{\tau_\mathrm{drag}}, ~~~~ i=1,2,\label{eqn:mom_2lev}
\\
\pp {\theta_i}t + \bm{u}_i\cdot\nabla\theta_i + \omega(\theta_2-\theta_1)  &= \dfrac{\theta_{iE}-\theta_i}{\tau_\mathrm{rad}}, ~~~~~~~~~~~~~~~~~ i=1,2,\label{eqn:thermo_2lev} 
\\
\dfrac{\Phi_2-\Phi_1}{\Pi_2-\Pi_1} &= -c_p \dfrac{\theta_1+\theta_2}{2},\label{eqn:hydros_2lev}
\\
\nabla \cdot \bm{u}_1 + 2\omega &= 0, \label{eqn:cont_2lev_1}
\\
\nabla \cdot \bm{u}_2 - 2\omega &= 0, \label{eqn:cont_2lev}
\end{align}
where the subscript 1 describes the upper layer, and 2 the lower layer. $\bm{u} = (u,v)$  is the horizontal velocity, $\omega$ the pressure velocity at the mid-level $p_0/2$ divided by $p_0$, $\Phi$ the geopotential height, and $\theta$ the potential temperature. $\Pi = (p/p_0)^{R/c_p}$ is the Exner function, where $R$ is the specific gas constant and $c_p$ the isobaric heat capacity (hereafter, $R/c_p = 2/7$, as for a diatomic gas). $f = 2\Omega \sin\phi$ is the Coriolis parameter, where $\Omega$ is the planetary rotation rate and $\phi$ is latitude. A Rayleigh drag, with timescale $\tau_\mathrm{drag}$, is applied in the lower layer. 

The thermodynamic forcing term consists of a relaxation towards a prescribed potential temperature profile $\theta_{iE}(\phi,\lambda)$, where $\lambda$ denotes longitude, on a time scale $\tau_\mathrm{rad}$. We use 
\begin{equation}\label{eqn:thetaE}
    \theta_{iE}(\phi,\lambda) = \left\{
    \begin{array}{ll}
        (\Delta\Theta_h - \Delta\Theta_v \ln\Pi_i)\;\cos\phi \;\max(0,\cos\lambda)& \text{for tidally-locked planets},\\
        (\Delta\Theta_h - \Delta\Theta_v \ln\Pi_i)\;\cos\phi \;\dfrac{1}{\pi} &\text{for non tidally-locked planets}.
    \end{array}\right.
\end{equation}
The meridional structure is taken proportional to that of the stellar irradiation in the absence of axial tilt (i.e., $\propto \cos\phi$). The same is true for the zonal structure on tidally-locked planets: it varies as $\cos\lambda$ on the day side and vanishes on the nightside.
The factor $1/\pi$ in the second expression of \eqref{eqn:thetaE} ensures that the mean $\theta_E$, a proxy for the stellar irradiation, is the same for tidally-locked and non-tidally-locked planets for given $\Delta\Theta_h$ and $\Delta\Theta_v$. 
Finally, the assumption of a zonally symmetric forcing for non tidally-locked planets is adequate when the thermal relaxation scale $\tau_\mathrm{rad}$ is much longer than the planet's rotation period $2\pi/\Omega$. This is not the case, for example, on Venus: there, the effect of thermal tides is of primary importance \citep[e.g][]{Takagi2007}.

In order to reduce the number of model parameters we nondimensionalize equations \eqref{eqn:mom_2lev}--\eqref{eqn:cont_2lev} following \cite{Potter2014}. Scaling length with $a$ (the planetary radius), time with $(2\Omega)^{-1}$, potential temperature with $\Delta\Theta_h$, geopotential with $c_p\Delta\Theta_h$, horizontal velocity with $c_p\Delta\Theta_h / (2\Omega a)$ (from geostrophic balance), and using a vertical velocity scale consistent with continuity, the governing equations become 
\begin{align}
\pp {\ubb_i} t + Ro_T\left(\bm{u}_i\cdot\nabla\bm{u}_i + \omega(\bm{u}_2 - \bm{u}_1) \right) + \hat f \bm{k}\times\bm{u}_i &= - \nabla \Phi_i - \delta_{i2}E \bm{u}_i, ~~~~ i=1,2,\label{eqn:mom_2lev_ndim}
\\
\pp {\theta_i}t +  Ro_T\left(\bm{u}_i\cdot\nabla\theta_i + \omega (\theta_2-\theta_1) \right) &= \dfrac{\theta_{iE}-\theta_i}{T_\mathrm{rad}}, ~~~~~~~~~~~~~~~ i=1,2,\label{eqn:thermo_2lev_ndim} 
\\
\Phi_2-\Phi_1 &= -\gamma(\theta_1+\theta_2),\label{eqn:hydros_2lev_ndim}
\\
\nabla \cdot \bm{u}_1 + 2\omega &= 0, \label{eqn:cont_2lev_1_ndim}
\\
\nabla \cdot \bm{u}_2 - 2\omega &= 0, \label{eqn:cont_2lev_ndim}
\end{align}
where all variables are now nondimensional, $\gamma = (\Pi_2 - \Pi_1)/2 \simeq 0.12$, and $\hat f = \sin\phi$. Three nondimensional control parameters appear: a thermal Rossby number, an Ekman number, and a nondimensional thermal relaxation time scale, given respectively by
\begin{equation}
    Ro_T = \dfrac{c_p\Delta\Theta_h}{(2\Omega a)^2},~~E = \dfrac{1}{2\Omega \tau_{\mathrm{drag}}},~~T_\mathrm{rad} = 2\Omega \tau_{\mathrm{rad}}.
\end{equation}
A fourth nondimensional parameter controls the vertical structure of $\theta_E$: $\mathcal{S} = \Delta \Theta_v/\Delta \Theta_h$. It is worth noting that in the framework of this paper, where the forcing temperature vanishes as the poles, $Ro_T$ is approximately the square of the ``weak temperature gradient parameter'' $\Lambda$ of \cite{Pierrehumbert2019}.

Numerical integrations on the sphere are performed using Dedalus \citep{Burns2020}, an open framework for solving partial differential equations using spectral methods. In addition to the fluid equations themselves, we use a small fourth-order horizontal hyperdiffusion in the momentum and thermodynamic equations to damp the enstrophy and tracer cascades at high wavenumbers, with values tuned to ensure numerical stability.

\section{Quasi-linear Processes Driving Superrotation} \label{sec:theory}

We now investigate the nature of the eddies that can drive superrotation. We do this by linearizing the model about basic states representative of tidally-locked planets and slow rotators and then examining the nature of the resulting eddy fluxes.

Denoting zonal averages with an overbar, the Eulerian-mean zonal momentum equation in a frictionless upper layer reads 
\begin{equation}\label{eqn:eddy_accel}
     \pp {\overline{u_1}} t = \overline{(\hat f+Ro_T\zeta_1)v_1} - Ro_T\;\overline{\omega(u_2-u_1)}.
\end{equation}
where overbars denote zonal averages, $\zeta_1$ is the relative vorticity of the upper layer, and we have used the identity $\bm{u}_1\cdot\nabla\bm{u}_1 = ({1}/2)\nabla(\bm{u}_1\cdot\bm{u}_1) + \zeta_1 \bm{k} \times \bm{u}_1$. For an equatorially symmetric circulation $v = 0$ at the equator, hence the only term that can positively accelerate the equatorial jet is the vertical momentum transport term in \eqref{eqn:eddy_accel}. 
Hence, superrotation requires $\overline{\omega(u_2-u_1)} < 0$ at the equator meaning that vertical motion generates a flux of eastward momentum from the lower to the upper layer. (A related condition for superrotation exists in $1.5$-layer shallow water models, but the absence of lower-layer flow ($u_2 = 0$) implies that the equatorial jet must be westward at some longitudes \citep{Showman2011}. No such requirement exists in the 2-level model.)
We now explore two classes of planetary-scale waves in tidally-locked planets and slow rotators that meet this condition.

\subsection{Tidally-locked planets} \label{subsec:theory_tl}
Tidally-locked planets are driven by a zonally varying thermal forcing. Because the cooling is near-uniform on the nightside, while the heating is much stronger near the equator on the dayside, there is also a zonal-mean net-heating gradient between the equator and the poles. Formally, one can expand the nondimensional version of the equilibrium potential temperature profile \eqref{eqn:thetaE} in Fourier series in longitude:
\begin{equation}
    \theta_{iE}(\phi,\lambda) = \left(\dfrac{1}{\pi} + \dfrac{1}{2}\cos\lambda + \dots\right)(1 - \mathcal{S} \ln\Pi_i)\;\cos\phi.
\end{equation}
One may then regard the total circulation as a sum of the response to the first term (an axisymmetric thermal forcing) and the second term (a wavenumber-one forcing), neglecting higher-order terms in the expansion. 

The axisymmetric part of the forcing inhibits superrotation, for two reasons. The first is that it leads to a thermally direct axisymmetric circulation, which exports angular momentum away from the equator. The second is that, by angular momentum conservation, this circulation leads to eastward midlatitude jets. These jets may  be baroclinically unstable generating Rossby waves that deposit westward momentum equatorward of their source region \citep{Vallis2017}.  

The linear response to the wavenumber-one part of the forcing is the well-known Matsuno--Gill circulation pattern\footnote{\cite{Matsuno1966} first studied the problem of the linear response of Earth's tropical atmosphere to a sinusoidal heat source with a particular meridional structure. \cite{Gill1980} extended this work to isolated heat sources with arbitrary meridional structures.}. \cite{Showman2010} discussed the shallow-water Matsuno--Gill problem in the context of tidally-locked planets, showing that the ensuing circulation leads to eastward momentum flux convergence at the equator when momentum fluxes from a lower quiescent layer are also present. The two-level model offers an enriched perspective: it does not require friction in the upper level, it allows for motion in the lower level, and the momentum transport between the two levels arises naturally in the equations of motion. 

The equations describing the Matsuno--Gill problem are obtained in the two-level system by linearizing \eqref{eqn:mom_2lev}--\eqref{eqn:cont_2lev} about a state of rest with uniform potential temperatures $\Theta_1$ and $\Theta_2$. We assume $\Theta_1-\Theta_2 = \mathcal{S}$; note that steady-state stratification in the nonlinear simulations of section \ref{sec:nonlinear} are between $\mathcal{S}$ and $4\mathcal{S}$.
\begin{subequations}
\begin{align}
    \hat f \bm{k}\times\bm{u}_1 + \nabla \Phi_1 &= 0,\label{eqn:mom_lin_1}\\ 
    \hat f \bm{k}\times\bm{u}_2 + \nabla \Phi_2 + E\bm{u}_2 &= 0,\label{eqn:mom_lin_2}\\
    - \mathcal{S}Ro_T\omega&= \dfrac{\theta_{1E}-\theta_1}{T_\mathrm{rad}},\label{eqn:thermo_lin_1}\\
    - \mathcal{S}Ro_T\omega&= \dfrac{\theta_{2E}-\theta_2}{T_\mathrm{rad}},\label{eqn:thermo_lin_2}
\end{align}
\end{subequations}
along with continuity and hydrostasy.  Here, $\theta_{iE} = \cos\phi\cos\lambda\,(1 - \mathcal{S} \ln\Pi_i)/2$ for $i=1,2$.

\begin{figure*}[!t]
\begin{center}
\plotone{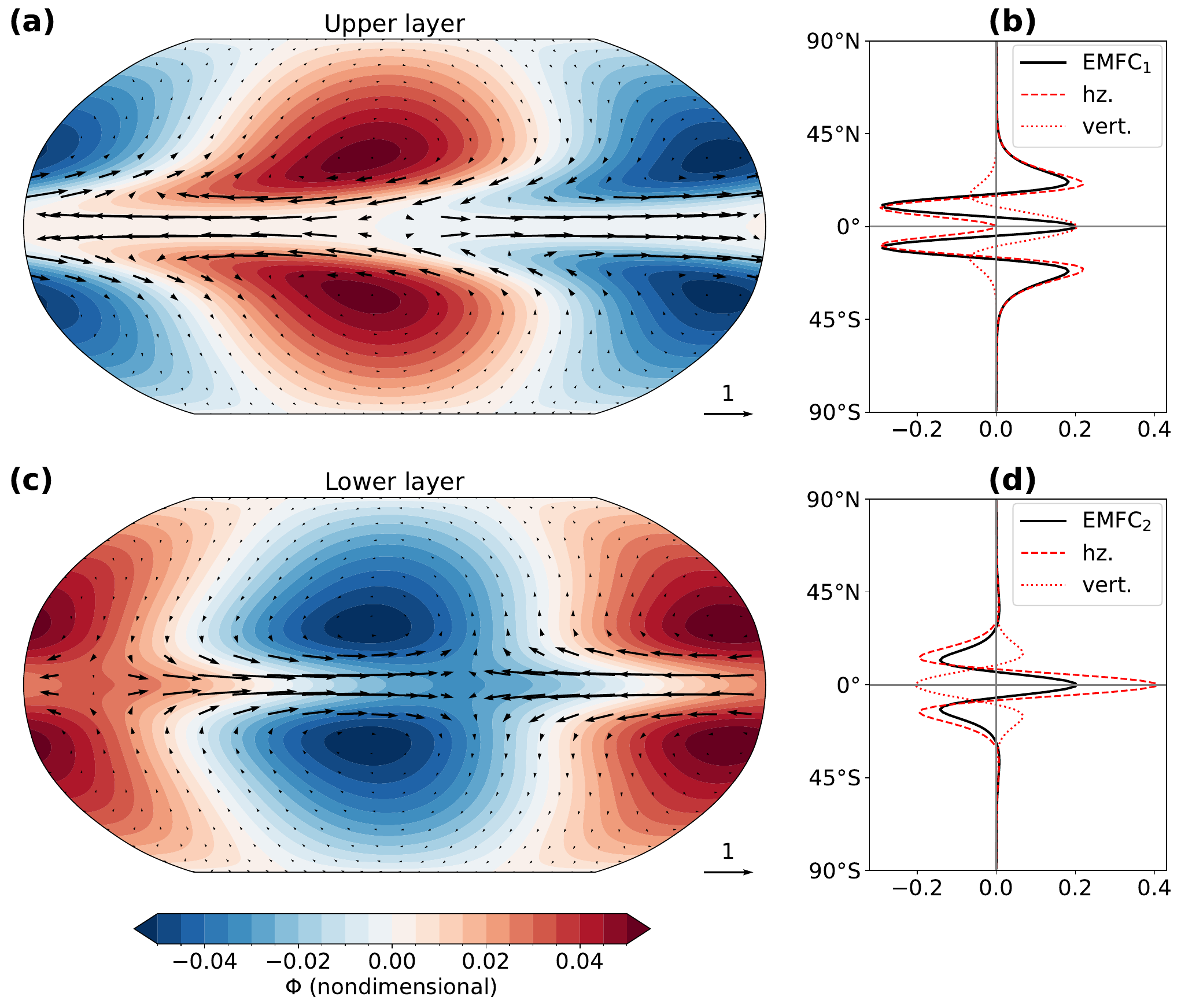}
 \caption{Solution of the Matsuno--Gill problem \eqref{eqn:mom_lin_1}--\eqref{eqn:thermo_lin_2} with $E=0.02$, $\mathcal{S} = 0.05$, and $Ro_T T_\mathrm{rad} = 20$, and its eddy momentum flux convergence. (a) Upper-layer geopotential $\Phi_1$ (shading) and wind $\bm{u}_1$ (arrows). (b) EMFC in the upper layer (solid), its horizontal convergence component (red dashed), and its vertical convergence component (red dotted) (see eq. \ref{eqn:EMFC1}). (c, d) As (a,b), except for the lower layer (see eq. \ref{eqn:emfc2}) } \label{fig:Gillpattern}
\end{center}
\end{figure*}

A sample solution of \eqref{eqn:mom_lin_1}--\eqref{eqn:thermo_lin_2}, solved on the sphere with $E=0.02$, $\mathcal{S} = 0.05$, and $Ro_T T_\mathrm{rad} = 20$, is shown in Fig. \ref{fig:Gillpattern}. The lower layer geopotential field (Fig. \ref{fig:Gillpattern}c) illustrates the classical structure of the response to equatorial heating, composed of an equatorial Kelvin wave and two off-equatorial Rossby waves which together form an eastward-pointing chevron pattern. The ensuing horizontal winds are such that $u_2v_2 < 0$ north of the equator, and $u_2v_2 > 0$ south of the equator. This means that $\partial_\phi(u_2v_2)<0$, which contributes to eastward eddy momentum flux convergence (hereafter EMFC): indeed, one can show that the total EMFC in the lower layer is 
\begin{equation}\label{eqn:emfc2}
    \mathrm{EMFC}_2 = Ro_T\left(\overline{\zeta_2v_2} - \overline{\omega(u_2-u_1)}\right) = - Ro_T \left(\dfrac{1}{\cos^2\phi} \pp {} \phi \left(\overline{u_2v_2} \cos^2\phi\right) - \overline{\omega(u_1+u_2)} \right),
\end{equation}
where the first term in the last expression represents the horizontal convergence of zonal momentum, and the second term is the vertical convergence of zonal momentum. Fig. \ref{fig:Gillpattern}d shows that the latter is negative in the lower layer, canceling about half of the momentum convergence by the horizontal flow. 

In the upper layer, the absence of drag mandates that the eddy geopotential field be uniform at the equator (Fig. \ref{fig:Gillpattern}a). Thus, there is no chevron-like pattern and the horizontal winds do not converge eastward momentum at the equator (Fig. \ref{fig:Gillpattern}b). However, the vertical convergence of zonal momentum is equal and opposite to that in the lower layer: indeed, 
\begin{equation}\label{eqn:EMFC1}
    \mathrm{EMFC}_1 = Ro_T\left(\overline{\zeta_1 v_1} - \overline{\omega(u_2-u_1)}\right) = - Ro_T \left( \dfrac{1}{\cos^2\phi} \pp {} \phi \left(\overline{u_1v_1} \cos^2\phi\right) + \overline{\omega(u_1+u_2)} \right),
\end{equation}
and the last term is opposite to that in \eqref{eqn:emfc2}.
This leads $\mathrm{EMFC}_1$ to be positive at the equator (Fig. \ref{fig:Gillpattern}b).

The two-layer Matsuno--Gill pattern thus does lead to eastward equatorial acceleration in both layers. To understand how the acceleration depends on the various control parameters, note that
the solution to \eqref{eqn:mom_lin_1}-\eqref{eqn:thermo_lin_2} mainly depends on two parameters: $E$ and $\mathcal{S}Ro_T T_\mathrm{rad}$ (as well as an additional very weak dependence on $\mathcal{S}$ through the magnitudes of $\theta_{1E}$ and $\theta_{2E}$). We begin by showing that in the absence of low-level drag ($E=0$), the equatorial EMFC vanishes. Combining the upper layer vorticity equation $\nabla\times$\eqref{eqn:mom_lin_1} with continuity gives 
\begin{equation}
    -2\hat f \omega + \beta v_1 = 0
\end{equation}
Taking a $\phi$ derivative, noting that $\partial_\phi \omega$ vanishes at the equator, and using continuity once again leads to $4\omega = - \partial_\lambda u_1$ (valid only at the equator), which implies 
\begin{equation}
    \overline{\omega u_1} = 0~ \text{ at the equator}
\end{equation}
When $E=0$, one similarly shows that $\overline{\omega u_2} = 0$. Recalling that the EMFC reduces to $- Ro_T\;\overline{\omega(u_2-u_1)}$ at the equator for an equatorially-symmetric circulation (eq. \ref{eqn:eddy_accel}), one finds that, in the absence of drag, the Matsuno--Gill pattern does not produce an equatorial EMFC.

\begin{figure*}[!t]
\begin{center}
 \plotone{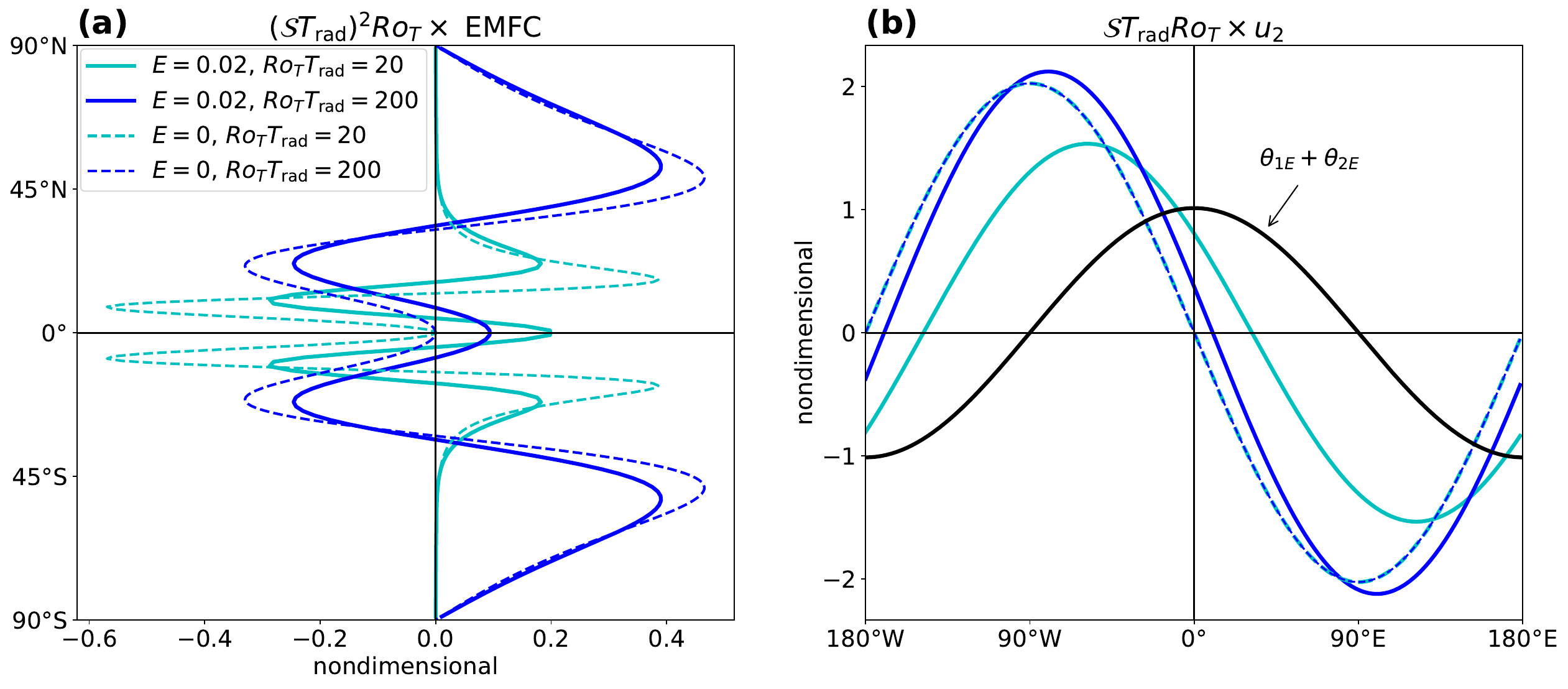}
 \caption{Gill model properties on the sphere with different input parameters. (a) Scaled upper-layer EMFC. (b) Scaled lower-layer zonal wind. The black line shows the vertically integrated reference potential temperature profile at the equator. Drag-free solutions (dashed) are obtained analytically in Appendix A. Solutions with low-level drag (solid) are obtained numerically.}\label{fig:gill_emfc_u2}
\end{center}
\end{figure*}

A full solution for the EMFC is obtained analytically for $E=0$ in Appendix A; we show that its magnitude scales as $Ro_T^{-1} \mathcal{S}^{-2} T_\mathrm{rad}^{-2}$ in the parameter regime explored in this paper. Fig. \ref{fig:gill_emfc_u2}a (dashed lines) shows the meridional pattern of EMFC obtained analytically for $E=0$ and two values of $Ro_T T_\mathrm{rad}$, scaled by $Ro_T \mathcal{S}^{2} T_\mathrm{rad}^{2}$. Both cases contain two off-equatorial maxima that shift equatorward as $Ro_T T_\mathrm{rad}$ is decreased. This poleward shift can be understood as follows: thermal relaxation is most effective at high latitudes, where potential temperature gradients can be balanced by the Coriolis force. A very effective thermal relaxation (small $Ro_T T_\mathrm{rad}$) yields small vertical motion at high latitudes (through \eqref{eqn:thermo_lin_1}--\eqref{eqn:thermo_lin_2}), hence weak meridional wind (through the vorticity balance) and weak zonal wind (through continuity). This leads to a weak EMFC at high latitudes for small $Ro_T T_\mathrm{rad}$.

For small $E>0$, the solutions (obtained numerically for $E=0.02$ and the same two values of $Ro_T T_\mathrm{rad}$) retain the same qualitative features (\figref{fig:gill_emfc_u2}a, solid lines). However, in both cases, the EMFC is positive at the equator. Why does the addition of drag enable eastward EMFC? We show in Appendix B that the upper-layer equatorial EMFC is 
\begin{equation}\label{eqn:VMT}
    \mathrm{EMFC}_1 = \dfrac{1}{\mathcal{S} T_\mathrm{rad}}\overline{(\theta_{1E}+\theta_{2E})u_2} ~\text{ at the equator.}
\end{equation}
In \figref{fig:gill_emfc_u2}b, we show patterns of $u_2(\lambda)$ at the equator for all the above solutions. Because the drag-free solutions of Appendix A scale as $(\mathcal{S} Ro_T T_\mathrm{rad})^{-1}$, we scale $u_2$ by $\mathcal{S} Ro_T T_\mathrm{rad}$. When $E=0$, $\theta_{1E}+\theta_{2E}$ (which acts as a proxy for vertical motion) and $u_2$ are exactly in quadrature, and $\overline{(\theta_{1E}+\theta_{2E})u_2} = 0$. When friction is included, the low-level wind pattern shifts eastward (solid lines in Figure \ref{fig:gill_emfc_u2}b), which allows for vertical motion to transfer eastward momentum to the upper layer. Interestingly, the shift is more pronounced as $Ro_T T_\mathrm{rad}$ is decreased. This explains why the equatorial peak in EMFC strengthens relative to its off-equatorial counterparts as $Ro_T T_\mathrm{rad}$ is decreased in Fig. \ref{fig:gill_emfc_u2}a.

The eastward shift in the low-level zonal wind can be qualitatively understood in the beta-plane shallow water Gill problem \citep{Vallis2017}.
As both thermal and mechanical damping are increased, the Kelvin wave cannot propagate as far to the east and so shifts westward. The converse happens to the Rossby component, which shifts eastward. At weak damping rates, the sum of their zonal wind fields shifts eastward for two reasons: the Rossby wave zonal wind field is about three times larger, and the Rossby wave feels a stronger damping (due to the wave's slower propagation speed), hence a given increase in damping shifts it further than the Kelvin wave. Decreased $Ro_T T_\mathrm{rad}$ is akin to an increase in thermal damping in the two-level model (eq. \ref{eqn:thermo_lin_1}-\ref{eqn:thermo_lin_2}). This explains the increase in the magnitude of the equatorial acceleration with decreasing $Ro_T T_\mathrm{rad}$, relative to the off-equatorial peaks.

In summary, the response to the zonally asymmetric heating that characterizes tidally-locked planets can be broken down in two main parts. The first one is an axisymmetric, thermally direct circulation. This part is expected to export eastward momentum away from equatorial regions. The second part is the response to the zonal wavenumber-1 component of the heating. When treated quasi-linearly, this response presents the following features:
\begin{itemize}
    \item In the upper layer, its EMFC pattern exhibits two off-equatorial maxima and one equatorial maximum.
    \item The magnitude of the off-equatorial maxima scales as $Ro_T^{-1} \mathcal{S}^{-2}T_\mathrm{rad}^{-2}$, in the parameter regime considered herein.
    \item The equatorial maximum is positive if and only if there is low-level drag ($E>0$). It vanishes when $E=0$.
    \item When $E>0$, the equatorial maximum strengthens relative to the off-equatorial maxima when $Ro_T T_\mathrm{rad}$ decreases.
\end{itemize}

For a given $Ro_T$, higher $T_\mathrm{rad}$ leads to weaker equatorial EMFC, and one may expect a circulation dominated by the midlatitude jets (spun up by either the thermally direct circulation or the eddy flow). These jets may accelerate the equatorial flow westward if they become baroclinically unstable and radiate Rossby waves breaking at low latitudes \citep[see, e.g,][]{Vallis2017}. Low $T_\mathrm{rad}$, on the contrary, leads to a strong and equatorially-focused eddy acceleration, and likely to superrotation. While this quasi-linear picture suggests that superrotation should be less favored at high-$Ro_T$, stronger nonlinearity may render it less relevant in that regime. Section \ref{sec:nonlinear} explores the equilibrated state of fully nonlinear simulations across a wide range of $Ro_T$ and $T_\mathrm{rad}$ to test these ideas.

\subsection{Slow rotators} 
      \label{subsec:theory_axi}
In our idealized picture, non-tidally-locked planets have an entirely zonally symmetric forcing and a solution in the form of a non-superrotating, axi-symmetric circulation exists. Thus, another mechanism is needed produce non-axisymmetric eddies that can accelerate the equatorial atmosphere, and indeed \citet{Iga2005} found that the  interaction of midlatitude Rossby waves with an equatorial Kelvin wave (the ``RK mode'') can give rise to an instability producing eastward EMFC. Although the pattern of this instability resembles in some ways the Matsuno-Gill pattern, the mode does not arise as a response to a stationary forcing and has a non-zero eastward propagation speed. \cite{Wang2014} extended the analysis using a primitive equation model, and found that an unstable RK mode can exist whenever the midlatitude jets Doppler-shift the Rossby wave phase speed to match that of the equatorial Kelvin wave. A physical proximity of the Rossby and Kelvin waves is also needed, otherwise the interaction is weak and zero if there is no overlap at all.

Being symmetric about the equator, the RK mode cannot converge momentum there in the absence of vertical momentum transport. Consequently, single-layer shallow water models cannot produce superrotation (by that mechanism) if that process is not parameterized \citep{ZuritaGotor2018}. However, RK modes that produce equatorial acceleration can be naturally captured in the 2-level model. To show this, we linearize \eqref{eqn:mom_2lev}--\eqref{eqn:cont_2lev} about a state of horizontally uniform potential temperatures $\Theta_1$ and $\Theta_2$ (with $\Theta_1-\Theta_2=\mathcal{S}$), with a barotropic background zonal wind $\bm{U} = (U(\phi),0)$, and without friction (the modes still appear in the presence of friction -- we are merely trying to show that friction is not a necessary component here). The equations read:
\begin{align}
    \partial_t\bm{u}_i + Ro_T\left(\bm{U}\cdot\nabla\bm{u}_i + \bm{u}_i\cdot\nabla\bm{U}\right) + \hat f \bm{k}\times\bm{u}_i + \nabla \Phi_i &= 0, ~~~ i=1,2,\label{eqn:mom_eigen}\\ 
    \partial_t\theta_i + Ro_T\left(\bm{U}\cdot\nabla\theta_i - \mathcal{S}\omega\right) + \dfrac{\theta_i}{T_\mathrm{rad}} &= 0, ~~~ i=1,2,\label{eqn:thermo_eigen}
\end{align}
along with continuity and hydrostasy. $\bm{u}_i$, $\theta_i$ and $\Phi_i$ are perturbation quantities. Following \cite{ZuritaGotor2018}, $U(\phi)$ has two broad midlatitude jets centered on a latitude $\phi_0$ (here taken as $50^\circ$). $U(\phi)$ is defined in terms of its vorticity $\zeta$; a parameter $\alpha$ governs the strength of the jets, with strong, angular-momentum-conserving jets for $\alpha=0$ and no wind for $\alpha=1$:
\begin{align}
    \zeta(\phi) &= \dfrac{1}{Ro_T}\left\{\begin{array}{cl} (\alpha -1) f,&|\phi|<\phi_0 \\ (\cos^{-2}\phi_0 - \alpha \tan^2{\phi_0}-1)\hat f,&|\phi|>\phi_0 \end{array}\right.\\
    U(\phi) &= -\dfrac{1}{\cos\phi}\int_0^\phi\zeta \cos\phi\,\d\phi.
\end{align}

The ability of the Rossby and Kelvin waves to phase-lock depends on two conditions: they must have equal phase speeds, and they must overlap meridionally. The phase speed of midlatitude Rossby waves is on the order of the background wind speed $U(\phi_0)$. It can be shown from \eqref{eqn:mom_eigen}-\eqref{eqn:thermo_eigen} along with continuity and hydrostatic balance that the phase speed of equatorial Kelvin waves is
\begin{equation}
c = \sqrt{\dfrac{\gamma \mathcal{S}}{2Ro_T}}.
\end{equation}
The two waves have equal angular phase speeds if the Froude number $Fr = [U(\phi_0) / \cos\phi_0]/c$ is of order unity \citep[the factor $1/\cos\phi_0$ transforms the Rossby wave's phase speed into an angular velocity]{Wang2014}. In practice, we will see that RK modes exist at values of $Fr$ larger than 1, as the Kelvin wave is able to propagate faster than $c$. The second condition, that of spatial overlap, may be measured as the ratio of the meridional extent of Kelvin waves (given by the equatorial Rossby radius $L_d = \sqrt{Ro_T c}$) to the Rossby wave latitude $\phi_0$ measured in radians. 

\color{black}
\begin{figure*}[!t]
\begin{center}
 \plotone{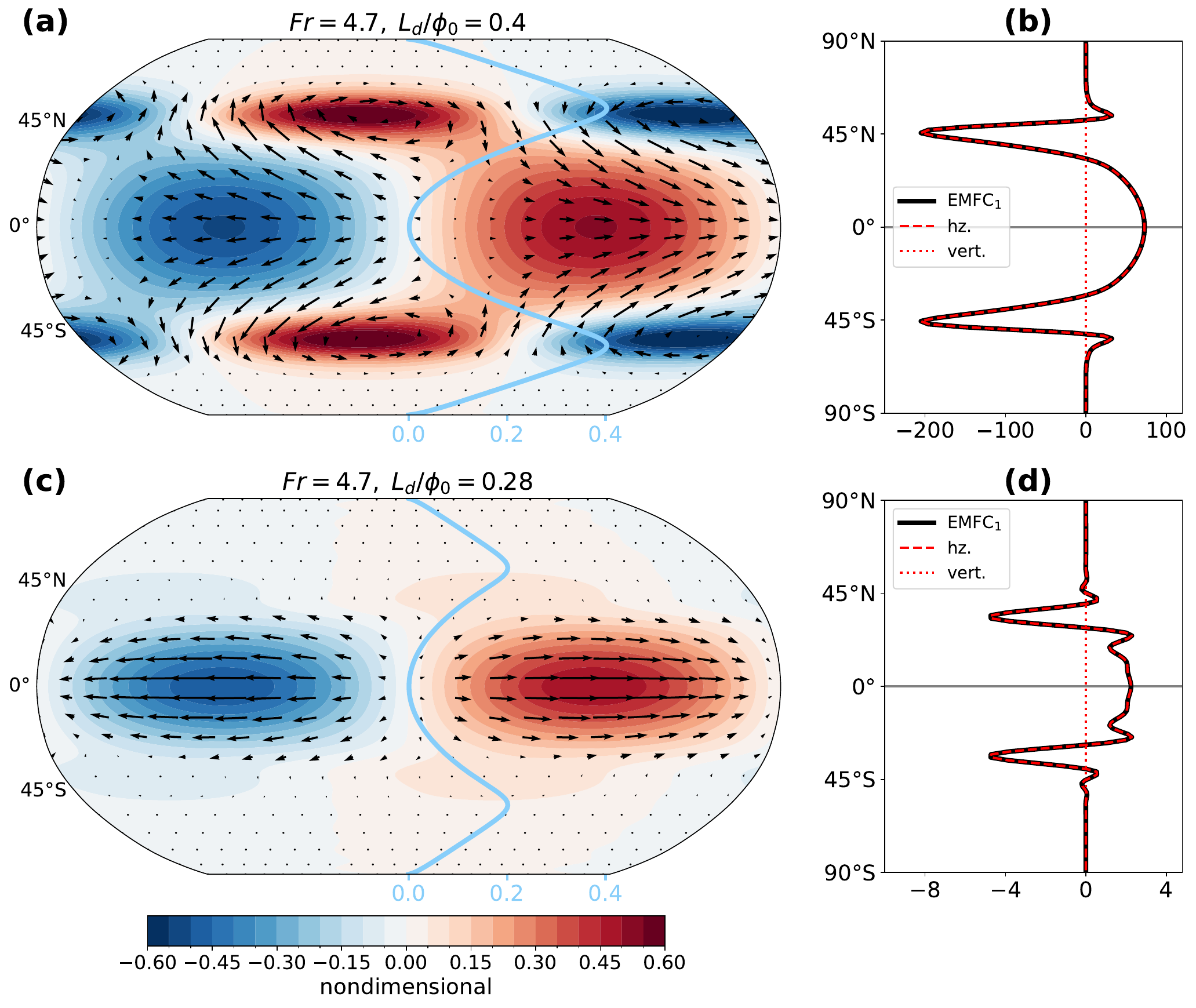}
 \caption{RK eigenmodes with $E=0$, $\mathcal{S} = 0.05$, and  $T_\mathrm{rad} = 200$. \textbf{(a)} Upper layer geopotential $\Phi_1$ (shading) and wind $\bm{u}_1$ (arrows), for $Ro_T = 10$ and $\alpha = 0$, corresponding to $Fr = 4.7$ and $L_d / \phi_0 = 0.4$. The thick blue line shows the local Rossby number, i.e. $Ro_T$ times the background wind profile (scale at the bottom of the panel). \textbf{(b)} Upper layer EMFC and its decomposition, as in Fig. \ref{fig:Gillpattern}b. \textbf{(c, d)} As \textbf{(a,b)}, except with $Ro_T = 2.5$ and $\alpha = 0.5$, corresponding to $Fr = 4.7$ and $L_d\phi_0 = 0.28$. The mode amplitudes are normalized by their mean upper layer kinetic energy.}\label{fig:eigenmodes}
\end{center}
\end{figure*}

We solve for zonal wavenumber-one eigenmodes of the system \eqref{eqn:mom_eigen}--\eqref{eqn:thermo_eigen} with two different sets of input parameters. Anticipating the simulation results of  \secref{subsec:nonlin_axi}, where the spatial overlap condition will prove more restrictive than that on $Fr$, we chose to vary the former and fix the latter.  We fix $T_\mathrm{rad} = 200$ and $\mathcal{S}=0.05$ and choose $Ro_T=10$, $\alpha = 0$ (a small / slowly rotating planet with angular-momentum conserving jets, that has Fr$ = 4.7$ and $L_d/\phi_0 = 0.4$) and $Ro_T=1.5$, $\alpha = 0.6$ (a larger / faster rotating planet with weaker jets, that has Fr$ = 4.7$ and $L_d/\phi_0 = 0.28$) as test cases.

In the first case (\figref{fig:eigenmodes}a), the most unstable mode has a typical RK pattern, with a broad equatorial Kelvin wave (due to the large deformation radius) coupling with Rossby waves propagating along the jets. The resulting eastward-pointing chevron pattern converges eastward momentum flux in a broad region from 30$^\circ$S to 30$^\circ$N (\figref{fig:eigenmodes}b). Because the vertical convergence of momentum fluxes $-Ro_T \overline{\omega(u_1+u_2)}$ vanishes (in this drag-free case, the symmetry of the governing equations imposes $\ubb_1+\ubb_2=0$), the total EMFC favors superrotation. The mode has an eigenfrequency of $0.23$, close to the theoretical frequency of the $n=1$ equatorial Kelvin wave, $Ro_T c = \sqrt{\gamma\mathcal{S}Ro_T/2}\simeq 0.17$. It has a large growth rate ($0.09$, about one over a rotation period).

An RK mode is also present in the second case (Figure \ref{fig:eigenmodes}c-d). The Rossby and Kelvin components do not interact as strongly, as the meridional scale of the Kelvin wave is smaller, resulting in a weak negative growth rate and nearly vanishing equatorial EMFC. Further investigation suggests that with the values of $T_\mathrm{rad}$, $\mathcal{S}$, and $\phi_0$ used in Figure \ref{fig:eigenmodes}, RK modes disappear below $L_d/\phi_0 \simeq 0.25$.

The dependence of the RK modes on $Fr$ is worth commenting on. Although one may expect that $Fr$ needs to be exactly 1 for Kelvin and Rossby waves to phase-lock, \cite{Wang2014} found unstable RK modes up to $Fr=4$. This requires either Rossby waves that are slower than the peak jet speed $U(\phi_0)$, or Kelvin waves that are faster than $c$. \cite{Wang2014} attributed their finding to slower Rossby waves. We found (not shown) that unstable RK modes may exist up to $Fr = 10$, and that the phase speed of the waves is weaker than the peak jet speed, but can be 2 to 3 times faster than $c$. This happens as the meridional flow (which is zero in a classical Kelvin wave but can be large for RK modes) strengthens convergence at the equator.

These results suggest that the two-level model contains the essential mechanisms to produce superrotation in a rational and tractable way in the presence of purely axisymmetric thermal forcing. In the next section we explore the effects of these mechanisms.

\section{Nonlinear Integrations over a Wide Range of Parameters} \label{sec:nonlinear}

In this section, we perform fully nonlinear integrations of the model, with both tidally-locked and axisymmetric thermal forcings and for a wide range of input parameters. Our goal is to test the qualitative, quasi-linear mechanisms presented in \secref{sec:theory}, including the behavior of superrotation as a function of $Ro_T$ and $T_\mathrm{rad}$ and the importance of surface friction for superrotation on tidally-locked planets, and to assess whether and how this simple framework can produce superrotation on slow rotators. The nonlinear 2-level model is envisioned as a bridge between simple models (e.g., linear and shallow water models) and GCMs, for it contains many of the processes of the latter (such as wave-mean-flow interactions, baroclinic instability and transient eddies) with the simplest possible vertical structure.

\subsection{Tidally-locked planets}
\label{subsec:2levTL}
All of the runs presented (except for the runs with increased friction and without friction at the end of this section) use $E=0.02$ and $\mathcal{S} = 0.05$. These are both relatively low values. $E$ is more representative of gaseous planets than rocky planets, and the low value of $\mathcal{S}$ represents relaxation towards a profile that is nearly neutral to convection. The presence of condensible species and/or stellar radiation absorbing species would likely increase this value. We perform a total of 121 runs, spanning two orders of magnitude in both $Ro_T$ (with values $0.1, 0.15, 0.25, 0.4, 0.65, 1.0, 1.5, 2.5, 4.0, 6.5, 10.0$) and $T_\mathrm{rad}$ (with values $10, 15, 25, 40, 65, 100, 150, 250, 400, 650, 1000$). Each run is integrated for 1000 rotation periods (i.e. until $t = 4000\pi$). A statistically steady state is reached after about 300 rotation periods, and steady-state values are obtained by averaging over the last 700.

\subsubsection{Equatorial jet strength and general features}\label{subsec:tl_nonlin_features}
\begin{figure*}[!t]
\begin{center}
 \plotone{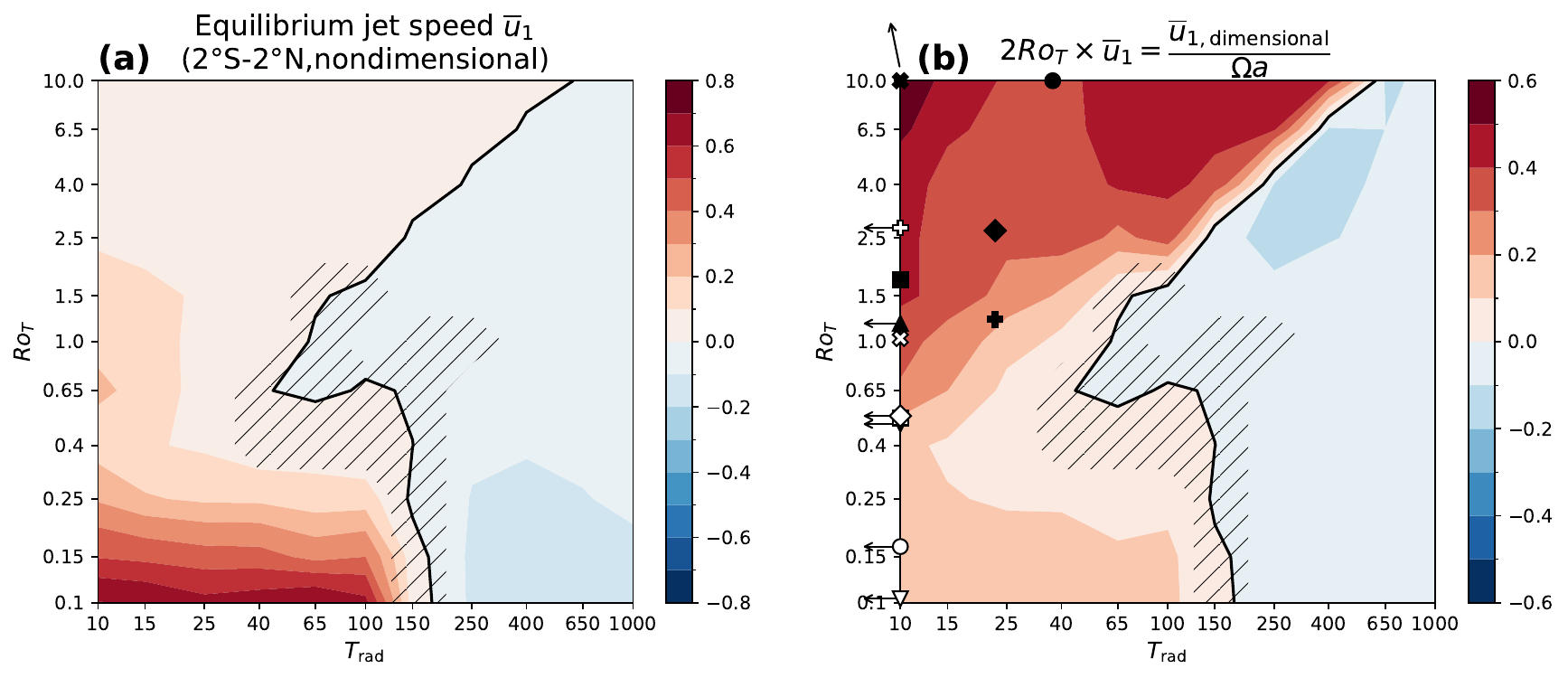}
 \caption{Equilibrium equatorial jet speed in tidally-locked planets. (a) Upper-level zonal-mean zonal wind speed $\overline{u_1}$ averaged 2$^\circ$S-2$^\circ$N, as a function of $Ro_T$ and $T_\mathrm{rad}$. (b) Same as (a), multiplied by $2Ro_T$. In both plots, the hatched region marks simulations for which $\overline{u_1}$ switches sign more than 10\% of the time in the last 700 rotation periods of the simulation. The thick black line marks the transition from superrotation to subrotation. In (b), black markers show approximate parameters for known tidally-locked terrestrial planets: GJ1132b ($\blacksquare$), LHS 1140 b ($\times$), Trappist 1b ($+$), Trappist 1c ($\blacklozenge$), Trappist 1d ($\bullet$), 55 Cancri e ($\blacktriangledown$), Kepler 10b ($\blacktriangle$). White-filled markers show approximate parameters for known hot Jupiters: HD 189733b ($\square$), HD 209458b ($\times$), HD 149026b ($+$), HAT-P-7b ($\lozenge$), WASP-18b ($\circ$), WASP-12b ($\triangledown$). Planets that fall outside of the regime diagram are brought to the nearest value; arrows are used to indicate planets for which $T_\mathrm{rad} < 5$ or $Ro_T > 20$. See Appendix C for details on the parameters and estimation.}\label{fig:regime_diagram_tl}
\end{center}
\end{figure*}  

\Figref{fig:regime_diagram_tl}a displays a regime diagram of the equilibrium equatorial jet speed in ($Ro_T$, $T_\mathrm{rad}$) space. For $Ro_T < 1$, the qualitative predictions from quasi-linear theory (Section \ref{subsec:theory_tl}) are verified: the jet speed decreases with both $Ro_T$ and $T_\mathrm{rad}$. For $Ro_T\geq 1$, these predictions break down: while the jet speed mostly decreases with $T_\mathrm{rad}$, it increases with $Ro_T$. It is not surprising that the Gill model loses quantitative accuracy when $Ro_T\geq 1$, as $Ro_T$ measures the importance of the nonlinear acceleration terms that it neglects. The behavior of the high $Ro_T$ regime is analyzed in further detail in the following. Another important point is that in this two-level framework, not all tidally-locked planets superrotate: with high enough $T_\mathrm{rad}$, subrotation appears for all the values of $Ro_T$ considered here. The transition between superrotation and subrotation does not happen abruptly in this ($Ro_T$, $T_\mathrm{rad}$) phase space: several simulations lying at the boundary between these states spend their time oscillating between the two (see hatched area in Figure \ref{fig:regime_diagram_tl}).

With the geostrophic scaling used in  \figref{fig:regime_diagram_tl}a, nondimensional wind speeds weaken with increasing $Ro_T$. Another way to nondimensionalize wind speeds would be to scale them by the planet's surface speed at the equator, $\Omega a$ (a factor of $2 Ro_T$ separates the former nondimensionalization from the latter). This scaling quantifies the excess angular momentum of the atmosphere relative to the planet's equatorial surface. Figure \ref{fig:regime_diagram_tl}b shows that superrotating jets at high $Ro_T$ generally exceed the planetary angular momentum by a larger fraction than at low $Ro_T$.

Various tidally-locked terrestrial planets and gas giants are featured in \figref{fig:regime_diagram_tl}b. The uncertainty of the positioning can be large, especially on $T_\mathrm{rad}$ (owing to poorly constrained atmospheric depth and composition - see Appendix C for details on the parameter estimation). However, the vast majority do lie in the low $T_\mathrm{rad}$ regime owing to the proximity to their host stars, leading to high equilibrium temperatures. As a consequence, all are localized in the superrotating regime.

\subsubsection{Low $Ro_T$ regime}

\begin{figure*}[!t]
\begin{center}
\includegraphics[width=0.9\textwidth]{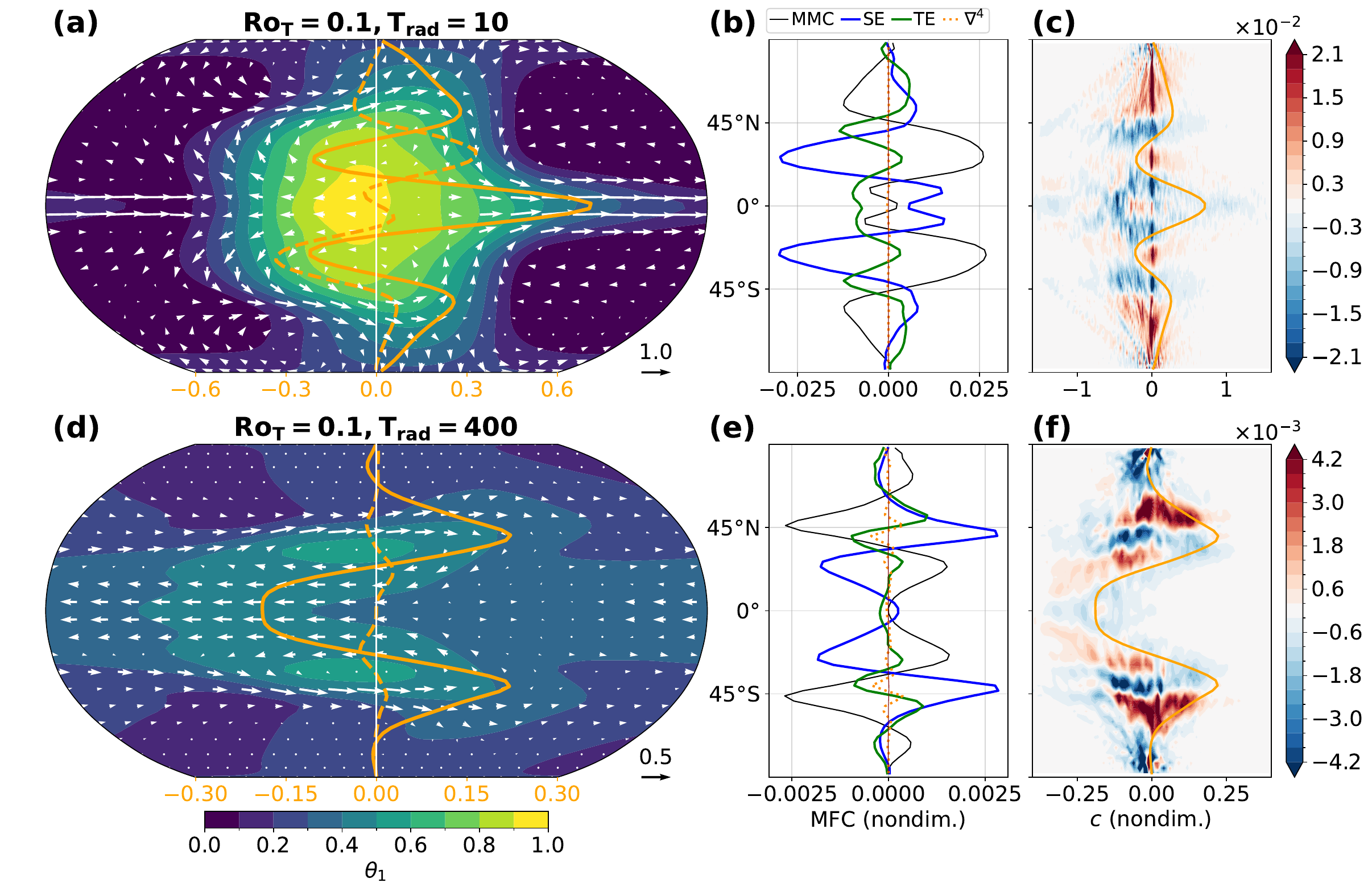}
 \caption{Equilibrium properties of two $Ro_T=0.1$ tidally-locked simulations, with $\textbf{(a,b,c)}$ $T_\mathrm{rad} = 10$ and $\textbf{(d,e,f)}$ $T_\mathrm{rad} = 400$. $\textbf{(a,d)}$ Mean upper level potential temperature (shading) and winds (arrows). Orange lines show $\overline{[u_1]}$ (solid) and $\overline{[v_1]}$ (dashed, multiplied by 5 for visibility), with a scale shown at the bottom of each axis. $\textbf{(b,e)}$ Upper-level zonal-mean zonal momentum budget \eqref{eqn:mmc_se_te} in equilibrium. Shown are the contributions of the mean meridional circulation (black), stationary eddies (blue), transient eddies (green), and hyperdiffusion (dotted orange). The small residual due to $[\partial_t\overline{u_1}]$ and numerical diffusion in the time-stepping scheme is shown as a thin red line. $\textbf{(c,f)}$ Cospectra of upper-level transient EMFC in (latitude, phase speed) space. Solid orange lines show $\overline{[u_1]}$; Rossby waves generated at high latitudes preferentially break and deposit westward momentum along these lines.}\label{fig:lowRoT}
\end{center}
\end{figure*}

Spanning two orders of magnitude in each of the two control parameters, our simulations sample very different regimes. We now investigate the behavior of four representative simulations, roughly sitting at each corner of the parameter space. We being with the low-$Ro_T$ regime, choosing one run with weak thermal inertia ($T_\mathrm{rad} = 10$, which superrotates) and one with large thermal inertia ($T_\mathrm{rad} = 400$, which subrotates).  \Figref{fig:lowRoT}a,d depicts the time-mean state of the upper level. 

The low-$Ro_T$, low-$T_\mathrm{rad}$ regime has a superrotating equatorial jet and two westerly midlatitude jets, with easterly flow in subtropical regions. The combination of low $Ro_T$ \citep[which limits temperature homogenization by gravity waves at low latitudes; see, e.g.,][]{Pierrehumbert2019} and low $T_\mathrm{rad}$ (which imposes strong relaxation towards the forcing temperature profile) yields large temperature gradients. The low-$Ro_T$, high-$T_\mathrm{rad}$ regime features a more homogenous temperature profile due to the larger thermal relaxation time. Its zonal-mean circulation is somewhat Earth-like, with a retrograde equatorial jet and two westerly midlatitude jets. Both simulations feature Hadley cells (Figure \ref{fig:lowRoT}a,d, dashed orange lines), with poleward flow extending to about 45° for $T_\mathrm{rad}=10$ and 35° for $T_\mathrm{rad}=400$. Weak reversed cells are present within 10° of the equator at low $T_\mathrm{rad}$.

What leads to these very different zonal-mean wind profiles? Our analysis of the Gill model (\secref{subsec:theory_tl}) suggested that on tidally-locked planets, equatorial EMFC driven by the stationary response to the day-night insolation gradient should decrease with $T_\mathrm{rad}$. To verify whether this behavior holds in the fully nonlinear simulations, we separate the contributions from the zonal-mean flow (MMC, for mean meridional circulation), stationary eddies (SE), and transient eddies (TE) to the zonal-mean zonal momentum budget of the upper layer. As before, zonal averages and deviations from these are denoted by $\overline{(\cdot)}$ and $(\cdot)'$. Time averages and deviations from these are denoted by $[\cdot]$ and $(\cdot)^\dagger$. \eqref{eqn:eddy_accel} is decomposed as
\begin{equation}\label{eqn:mmc_se_te}
\begin{array}{cl}
   \overline{\left[\dfrac{\partial u_1}{\partial t}\right]}
   =& \underbrace{ ( \hat f + Ro_T\overline{\left[\zeta_1\right]} ) \overline{[v_1]} - Ro_T\;\overline{[\omega]}\,\overline{[u_2-u_1]}}_\mathrm{MMC}\\
   &+ \underbrace{Ro_T\left(\overline{[\zeta_1]' [v_1]'} - \overline{[\omega]' [u_2-u_1]')}\right)}_\mathrm{SE}\\
   &+ \underbrace{Ro_T\left(\overline{\left[\zeta_1^\dagger v_1^\dagger\right]} - \overline{\left[\omega^\dagger (u_2-u_1)^\dagger\right]}\right)}_\mathrm{TE}.\\
\end{array}
\end{equation}
The left-hand side vanishes in equilibrium. The Gill model is relevant to the behavior of the SE term.

Equation \eqref{eqn:mmc_se_te} is evaluated in statistical equilibrium in \figref{fig:lowRoT}b,e. In this low $Ro_T$ regime, the meridional pattern of SE momentum flux convergence qualitatively resembles that of the Gill model (Figure \ref{fig:gill_emfc_u2}a), with three local maxima (two in midlatitudes and one at the equator). The overall magnitude of the SE term and the strength of its equatorial maximum relative to the midlatitude peaks both decrease with $T_\mathrm{rad}$, in line with the Gill model. Interestingly, in both cases, the SE term is mostly balanced by transient eddies close to the equator, with the MMC having a weak contribution there. Similar behavior was previously observed in a GCM with Earth-like parameters \citep{Lutsko2018}.

To get more insight into the nature of the westward momentum flux at low latitudes, we spectrally decompose of the transient EMFC \citep[][see Appendix D]{Mitchell2010,Randel1991}. \Figref{fig:lowRoT}c,f shows the contribution of eddies of a given phase speed to the TE term at each latitude. Eddies generated at midlatitudes will tend to break and deposit their momentum along their critical line, i.e., where their phase speed equals the zonal-mean zonal wind $\overline{u_1}$. In the subrotating case (Fig. \ref{fig:lowRoT}f), high latitude eddies converge eastward momentum slightly poleward of the location of the jets, and diverge it along the critical line equatorward of it \citep{Randel1991}. Close to the equator, eddies of various phase speeds deposit westward momentum, but these are not clearly connected to breaking high-latitude waves. In the superrotating case, the pattern of EMFC around the midlatitude jets is relatively similar (with westward momentum deposition equatorward of the jet), but the behavior of low latitudes is very different, with contributions from eddies of much larger phase speeds.

In summary, the low-$Ro_T$ tidally-locked runs can broadly be classified into two categories: a superrotating one at low $T_\mathrm{rad}$, and a subrotating one at high $T_\mathrm{rad}$ (the boundary being at $T_\mathrm{rad}\simeq 150$, see Fig. \ref{fig:regime_diagram_tl}). Momentum flux convergence by stationary eddies behaves in a qualitatively similar way to the Gill model, providing increasingly weaker eastward equatorial acceleration as $T_\mathrm{rad}$ increases, and is balanced there by westward acceleration due to transient eddies. The Hadley circulation has a weak contribution to the zonal momentum balance in the equatorial zone.

\subsubsection{High $Ro_T$ regime}

\begin{figure*}[!t]
\begin{center}
\includegraphics[width=0.9\textwidth]{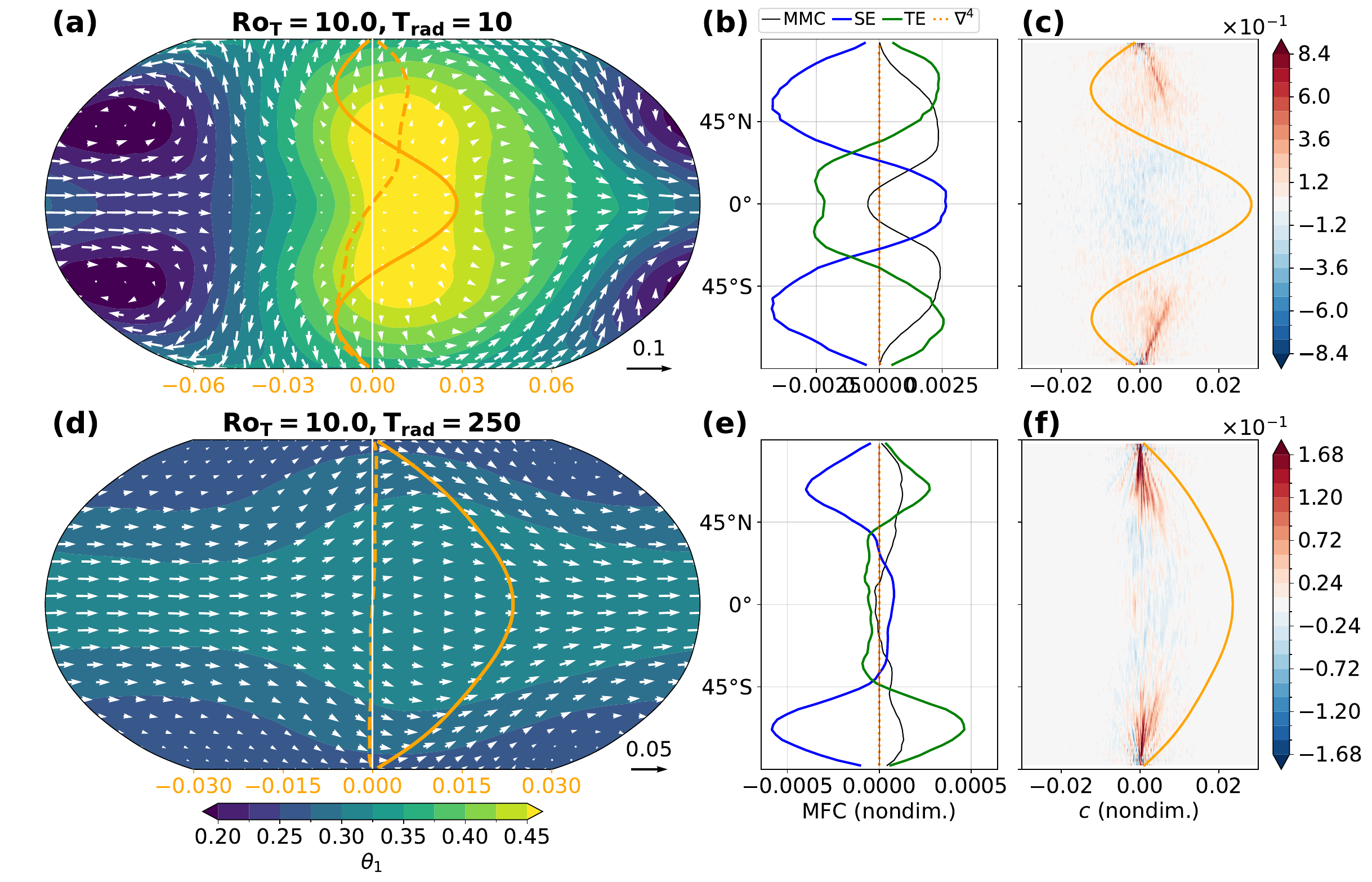}
 \caption{As in Fig. \ref{fig:lowRoT}, for two simulations with $Ro_T=10$ and $T_\mathrm{rad} = 10$ and 250.}\label{fig:highRoT}
\end{center}
\end{figure*}

The other end of the parameter space is illustrated in Fig. \ref{fig:highRoT} with two $Ro_T=10$ simulations. We choose two superrotating cases, with low and high $T_\mathrm{rad}$. Both cases have much more homegenous temperature distributions than their low $Ro_T$ counterparts (notice the difference in color scales between Fig. \ref{fig:lowRoT}a,d and Fig. \ref{fig:highRoT}a,d), due to efficient temperature smoothing by gravity waves. The $T_\mathrm{rad}=10$ case features a broad superrotating equatorial jet and easterly flow at high latitudes. The $T_\mathrm{rad}=250$ case has planetary-wide eastward flow. This configuration appears in GCMs but cannot occur in $1.5$-layer shallow water models, as some local retrograde flow is required at the equator to allow for vertical transport of eastward momentum in these models (as discussed in the introduction). Hadley cells extend all the way to the poles in both simulations (Fig. \ref{fig:lowRoT}a,d, dashed orange lines).

The zonal-mean zonal momentum budget \eqref{eqn:mmc_se_te} is shown in \figref{fig:highRoT}b,e. The Hadley circulation decelerates the equatorial flow, as its ascending branch imports air with lower angular momentum from the lower level. However, this contribution remains weaker than that of stationary and transient eddies. In this high-$Ro_T$ regime, the SE momentum flux convergence patterns strongly differ from the Gill model (there is only one maximum, at the equator, converging eastward momentum in a broad region). This is because the Gill model assumes no mean flow, while the mean flow strongly influences the shape of stationary eddies at high $Ro_T$ \citep{Hammond2018, Pierrehumbert2019}.

As in the low-$Ro_T$ regime, most of the SE term is balanced by transient eddies. This is somewhat surprising in light of the findings of \cite{Hammond2020}, which showed that in GCM simulations at high $Ro_T$ , the SE term was balanced by the vertical part of the MMC term, with transient eddies being weak. The two-level simulations presented here exhibit stronger time-variability than classical GCMs: while \cite{Komacek2020} reported time variability in global mean temperature and wind speed of around, respectively, 0.1--1\% and 1--10\%, these figures are 0.3--15\% and 10--40\% for $Ro_T\in[1,10]$ and $T_\mathrm{rad}\in[10,1000]$ in the two level model. This difference may arise from the strong vertical truncation inherent to the two-level model or the weak stratification of our relaxation temperature profile, and will be explored in a further study. 

Eastward EMFC at high latitudes mostly results from eddies of weak positive phase speeds. A broad spectrum of waves contributes to westward EMFC at low latitudes, although isolated signals of eastward acceleration are detectable (in the $Ro_T=250$ case, the most prominent one is related to the MRG wave dicussed in section \ref{sec:transition}).

\subsubsection{Propagating Rossby--Kelvin modes contribute to superrotation at high $Ro_T$}

Why does the propensity to superrotate increase at high $Ro_T$ (Fig. \ref{fig:regime_diagram_tl}), while the Gill model seemed to suggest the production of eddies with the opposite effect? In this section, we show evidence that RK modes may provide an answer to this question even in tidally-locked planets. For brevity, we choose the $Ro_T = 10$, $T_\mathrm{rad} = 250$ run analyzed above. However, several simulations (mostly those lying close to the transition between superrotation and subrotation, for $Ro_T>1$) exhibit similar behavior.

\begin{figure}[!t]
\begin{center}
 \plotone{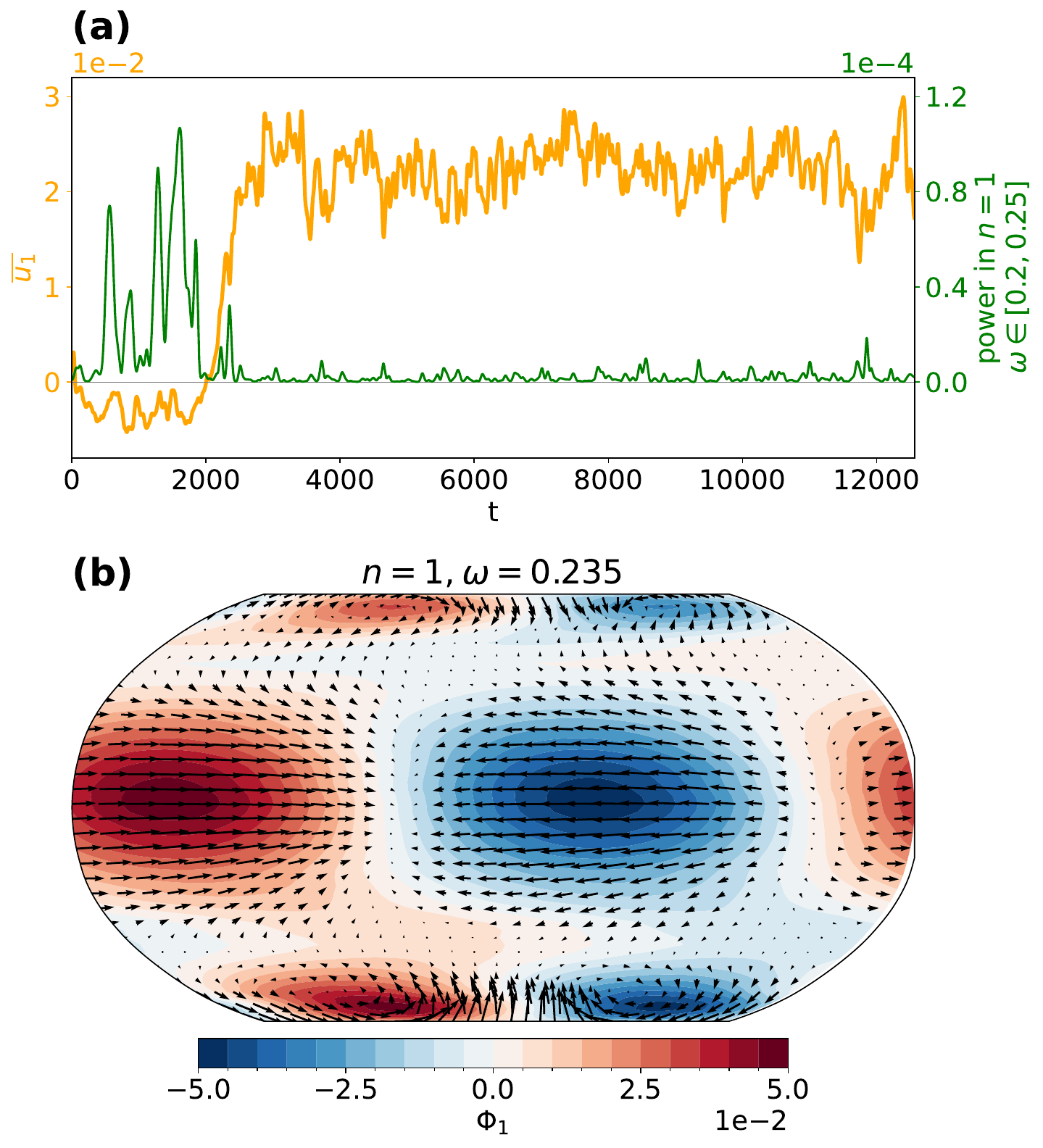}
 \caption{RK mode in a high-$Ro_T$ tidally-locked simulation ($Ro_T=10$, $T_\mathrm{rad} = 250$). (a) Timeseries of equatorial RK mode power, as measured as the power contained in the $[0.2,0.25]$ frequency band of the $n=1$ equatorial zonal wind (green line), along with timeseries of the equatorial zonal-mean zonal wind (orange line). Both timeseries are smoothed with a Gaussian filter with standard deviation $4\pi$ (1 rotation period). (b) Upper-layer geopotential and winds filtered for $n=1$ and $\omega = \omega_{\max}$, where $\omega_{\max}$ is the frequency that maximizes $K_{1,\omega}$ averaged over 30$^\circ$S - 30$^\circ$N.}\label{fig:RK_tidally_locked}
\end{center}
\end{figure}

We start by quantifying the presence of $n=1$ waves within the RK wave frequency band at the equator. More precisely, we form a timeseries by taking the $n=1$ Fourier component of the 30$^\circ$S - 30$^\circ$N averaged zonal wind. We perform a continuous wavelet transform \citep{Torrence1998} on this timeseries\footnote{We use Morlet wavelets with frequencies of the form $2^{-j/12} / (2\pi)$, where $j$ is a nonnegative integer.}, and average the power for $\omega\in[0.2,0.25]$ (motivated by the results of section \ref{subsec:theory_axi}). The result is shown as green lines in Fig. \ref{fig:RK_tidally_locked}a, alongside a timeseries of the equatorial zonal-mean zonal wind.

The simulation switches from subrotation to superrotation around $t = 2000$, with relatively stable zonal-mean equatorial winds before and after the regime change. The switch is preceded by a burst in wavenumber-1 equatorial wave activity in the RK frequency band. This activity mostly disappears once superrotation has emerged, although a weak burst around $t = 12000$ coincides with a temporary weakening of the equatorial winds. This behavior was observed in a GCM by \citet{ZuritaGotor2022} and is consistent with the fact that the RK mode requires weak equatorial winds for the coupling between the equatorial Kelvin wave and the midlatitude Rossby waves to occur.

We evaluate the spatial structure of this signal by filtering upper-layer geopotential and winds for $t\in[0,3000]$ for zonal wavenumber $n=1$ and $\omega = 0.235$ (the frequency that has the strongest contribution to eastward EMFC at the equator for $n=1$). The resulting pattern (Fig. \ref{fig:RK_tidally_locked}b) features a clear coupling between a low-latitude Kelvin wave and high-latitude Rossby waves. Although their phase difference is unlike the theoretical patterns of Fig. \ref{fig:eigenmodes}, they still combine in a way to converge eastward momentum at low latitudes.

This result strongly suggests that RK instability may drive the transition to superrotation even in some classes of tidally-locked planets, specifically for $Ro_T \gtrsim 1$. The presence of the instability may be favored by the weakness of the Gill response in that regime (section \ref{subsec:theory_tl}), as well as the relative axial symmetry of the basic-state flow (Fig. \ref{fig:highRoT}d).

\subsubsection{Importance of low-level drag}
One important prediction from the quasi-linear solutions of section \ref{subsec:theory_tl} is that the Gill pattern does not accelerate superrotation in the absence of low-level drag. To test whether this affects the appearance of superrotation in fully nonlinear runs, we re-run three simulations with $E=0$ (i.e., we switch off Rayleigh drag altogether). We choose to fix $T_\mathrm{rad} = 40$ (a value at which all simulations superrotate with our reference value of $E=0.02$), and use three values of $Ro_T$: $0.1$, 1, and 10.

\begin{figure*}
\begin{center}
  \includegraphics[width=\textwidth]{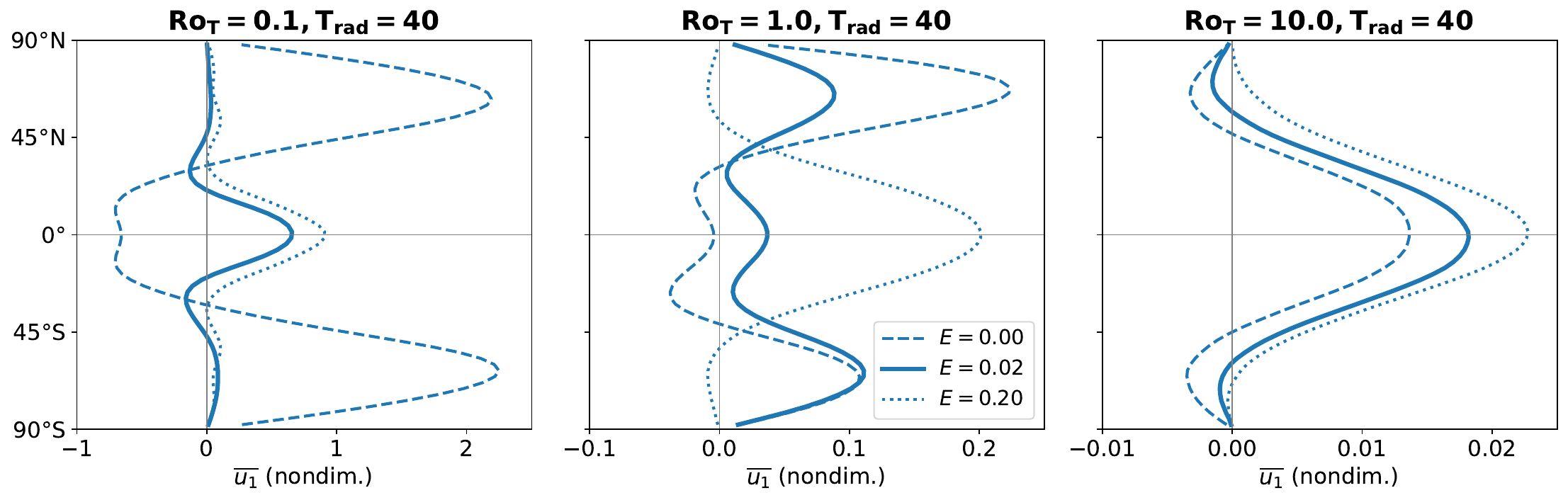}
 \caption{Upper level zonal-mean zonal wind in two-level runs with no drag (dashed lines), our reference drag value (solid lines, corresponding to damping time of 4 rotation periods), and strong drag (dotted lines, corresponding to a damping time of 0.4 rotation periods).}
 \label{fig:dragvnodrag}
\end{center}
\end{figure*}

The equilibrium $\overline{u_1}$ is shown in Fig. \ref{fig:dragvnodrag} for our control runs (solid) and no-drag runs (dashed). Shutting off low-level drag removes superrotation in the $Ro_T=0.1$ and $Ro_T=1$ runs, consistent with the Gill model behavior. In the $Ro_T=10$ case, however, the superrotating jet is only slightly weakened by the removal of drag. This confirms our previous conclusion that the Gill model is more relevant at low $Ro_T$. It is also consistent with our finding that RK instability is linked to the appearance of superrotation at high $Ro_T$: indeed, RK modes do not require surface drag to produce equatorial EMFC (section \ref{subsec:theory_axi}).

The comparison with shallow water models deserves a comment. $1.5$-layer shallow water models such as in \cite{Showman2011} prescribe drag in the upper layer, and have essentially infinite drag in the lower layer. Thus, ``drag-free'' solutions of such models are not comparable to the present two-level runs, because they are only drag-free in the upper layer (as are all runs presented in the present work) -- they do, in fact, superrotate. 

We also test the sensitivity to increased low-level drag, and run the same three simulations with $E=0.2$. This value is 10 times higher than our reference choice, and corresponds to a timescale of 0.4 rotation periods; it is thus on the upper end of drag values typically used in GCM studies of tidally-locked planets \citep{Liu2013}. Stronger drag results in stronger superrotating jets in all three cases, with a clear regime shift in the $Ro_T=1$ case from 3 jets to a single equatorial jet. This is consistent with the Gill pattern still being relevant in the appearance of superrotation at $Ro_T=1$.

\subsection{Axisymmetrically-forced planets} 
\label{subsec:nonlin_axi}

We now move away from the tidally-locked configuration, towards axisymmetric thermal forcing.  We use the same setup and set of parameters as in \ref{subsec:2levTL}, except for the equilibrium potential temperature profiles, which are zonally symmetric --- see \eqref{eqn:thetaE}.  Each run is integrated for 500 rotation periods (ample to reach statistical equilibration), with averaged values taken over the last 350 periods.
Our main goal is to isolate the mechanism of any superrotation and to see if and how it is tied to RK instability. 

\begin{figure*}[!t]
\begin{center}
 \plotone{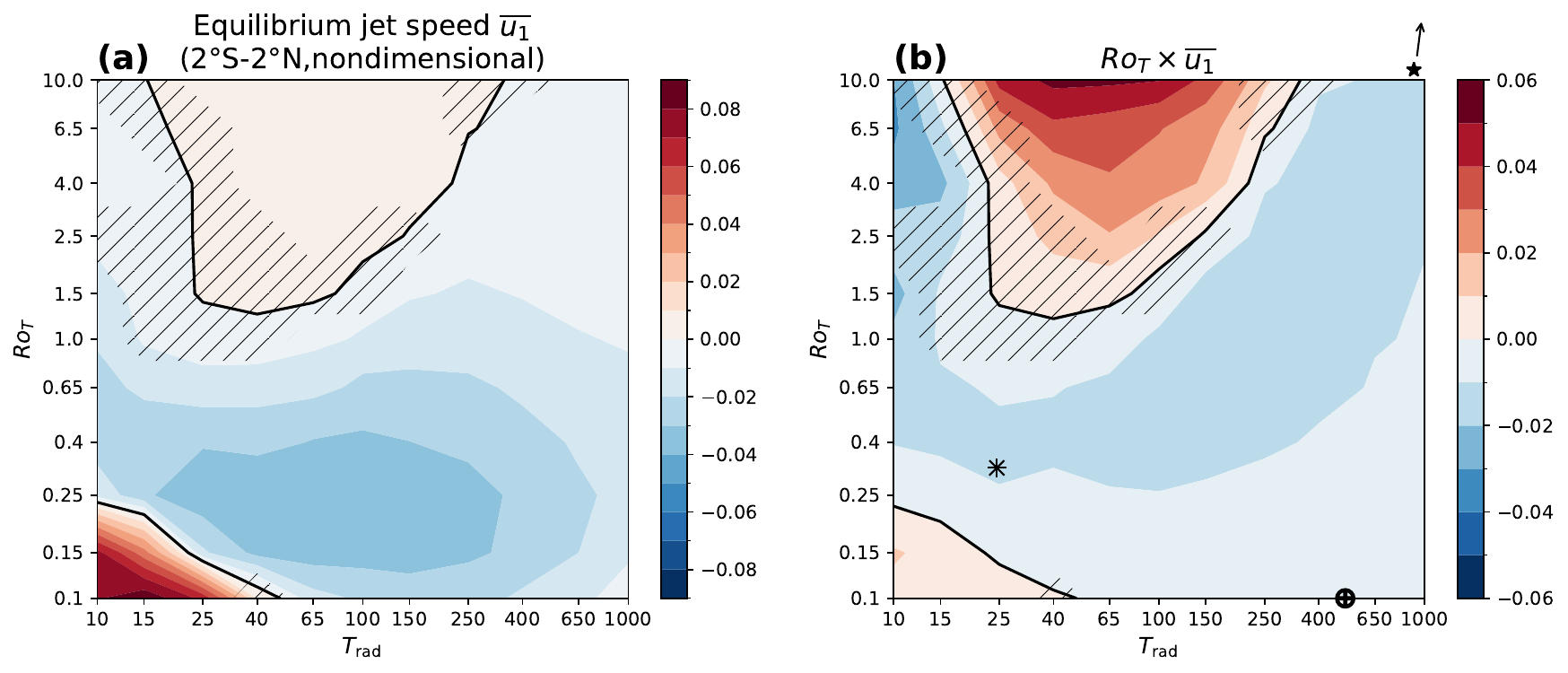}
 \caption{Same as Fig. \ref{fig:regime_diagram_tl}, for axisymmetrically-forced planets. The locations of Earth ($\oplus$) and Mars ($*$) are indicated; Titan ($\star$) lies outside of the regime diagram (see Appendix C).}\label{fig:regime_diagram_axi}
\end{center}
\end{figure*}
\Figref{fig:regime_diagram_axi}a displays a regime diagram of the equilibrium equatorial jet speed in the upper level, in ($Ro_T$, $T_\mathrm{rad}$) space. Most of these axisymmetrically-forced planets subrotate. Consistent with previous studies \citep[e.g.][]{Potter2014}, superrotation is favored at high-$Ro_T$ (here, $Ro_T\gtrsim 1.5$), albeit only when $T_\mathrm{rad}$ is within certain bounds. The $Ro_T$ threshold for superrotation is similar to previous studies \citep{Mitchell2010, Potter2014}, and the low-$T_\mathrm{rad}$ cutoff was previously observed in a GCM by \cite{DiasPinto2014}. As in the tidally-locked cases, the boundary between subrotation and superrotation is not well defined in all parts of the parameter space: some simulations, especially at high $Ro_T$ and low $T_\mathrm{rad}$, oscillate between both states even in statistical equilibrium. Somewhat surprisingly, a second region of the phase space (specifically at low $Ro_T$ and low $T_\mathrm{rad}$) superrotates. While we do not focus on this case, preliminary examination suggests that high-wavenumber mixed Rossby-gravity (MRG) waves accelerate superrotation in that regime. Finally, as in the tidally-locked case, the equatorial jet speed strengthens with $Ro_T$ when measured as a fraction of the planetary angular momentum (Fig. \ref{fig:regime_diagram_axi}b). Earth and Mars are clearly in the subrotating regime of this diagram. Titan has both high $Ro_T$ and high $T_\mathrm{rad}$ (Appendix C) and thus lies outside of the parameter range explored here.

\begin{figure*}[!t]
\begin{center}
 \includegraphics[width=\textwidth]{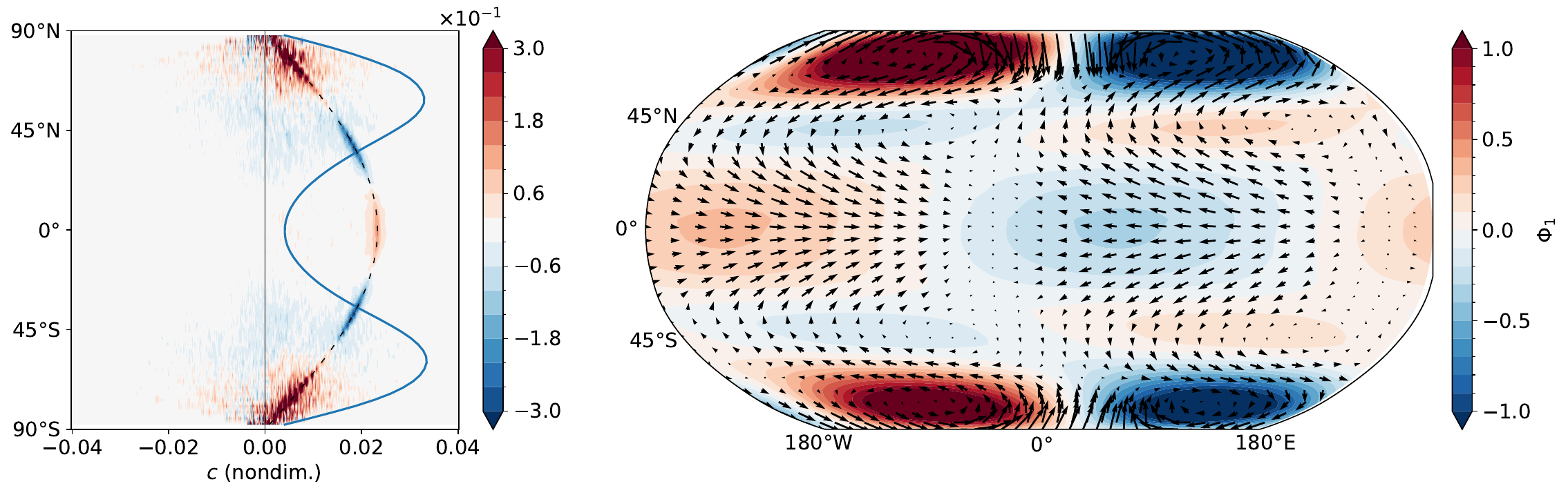}
 \caption{Evidence that RK modes accelerate superrotation in a nonlinear axisymetrically forced run with $Ro_T=10$ and $T_\mathrm{rad} = 25$. (Left) Cospectrum of upper-level EMFC in (latitude, phase speed) space. The solid blue line shows the upper-level zonal-mean zonal wind. The dashed black line shows the phase speed associated with a constant frequency $\omega = 0.23$ and wavenumber $n=1$. (Right) Upper-level zonal geopotential (shading) and zonal wind (arrows) filtered for $n=1$ and $\omega = 0.23$.}\label{fig:RK_axi}
\end{center}
\end{figure*}
Can the superrotation be linked to RK instability at high $Ro_T$? We provide evidence for this on one example ($Ro_T = 10, T_\mathrm{rad} = 25$), although all superrotating simulations behave in a similar fashion. The cospectrum of EMFC in the upper layer (Fig. \ref{fig:RK_axi}a) shows a clear signal at all latitudes corresponding to a frequency of $0.23$ and wavenumber $n=1$. This wave contributes most of the eastward acceleration at the equator. The upper layer geopotential and winds are filtered for this exact frequency and wavenumber (Fig. \ref{fig:RK_axi}b), and shows a classical RK pattern, similar to Fig. \ref{fig:eigenmodes}a, flanked by a set of high-latitude Rossby waves close to each pole. While the case presented in \figref{fig:RK_axi} has relatively simple behavior, the cospectra of other superrotating simulations can have more complex structures (not shown). The footprint of RK modes is present in all of them, but other types of waves (particularly a $n=2$ westward-traveling MRG wave) contribute to superrotation in some cases.  

\begin{figure*}[!t]
\begin{center}
 \includegraphics[width = \textwidth]{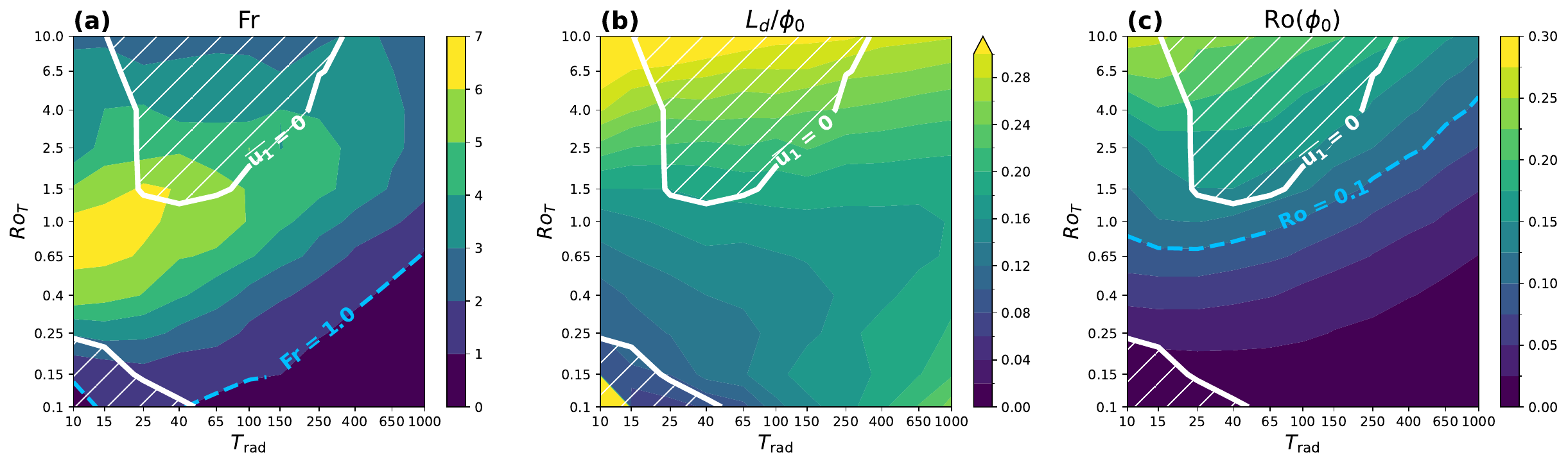}
 \caption{Diagnostics for the existence of RK instability in the suite of axisymmetrically-forced simulations. \textbf{(a)} Froude number.  \textbf{(b)} Ratio of the equatorial Rossby deformation radius to the latitude of maximum zonal wind speed. \textbf{(c)} Rossby number at the latitude of maximum zonal wind speed. Dotted blue lines in (a) and (c) indicate the lower bounds for RK instability from \protect{\cite{Wang2014}}. The hatched regions enclosed by solid white contours delineate the regions of the ($Ro_T$, $T_\mathrm{rad}$) phase space that superrotate.}\label{fig:Fr_Ro}
\end{center}
\end{figure*}

As explained in \secref{subsec:theory_axi}, the existence of unstable RK modes requires two conditions: a Froude number $Fr \geq 1$ and sufficient physical overlap between the waves, as measured by the ratio $L_d/\phi_0$. The values of $Fr$ and $L_d/\phi_0$ are shown in the parameter space of the simulations in Fig. \ref{fig:Fr_Ro}a. A basic-state wind $U(\phi)$ is obtained by averaging $(u_1 + u_2)/2$ over both hemispheres, and $\phi_0$ is taken as the latitude where $U$ maximizes. $Fr$ is calculated as $Fr = [U(\phi_0) / \cos\phi_0]/(U_\mathrm{eq} + c)$ to take into account Doppler-shifting of the equatorial Kelvin wave by the mean flow, with $U_\mathrm{eq}$ averaged between 30$^\circ$S and 30$^\circ$N. The stratification in $c$ is taken as the difference $\theta_1-\theta_2$ averaged over the same equatorial band. $Fr$ turns out to be larger than 1 over much of the parameter space, indicating that the midlatitude jets are fast enough for Rossby waves to phase-lock with equatorial Kelvin waves, except at low $Ro_T$ and high $T_\mathrm{rad}$ (e.g., for Earth-like planets).

Fig. \ref{fig:Fr_Ro}b shows that the spatial overlap between Rossby and Kelvin waves increases with $Ro_T$. Except for the low-$Ro_T$, low-$T_\mathrm{rad}$ regime, superrotation appears for $L_d/\phi_0 \gtrsim 0.18$ (a slightly lower value than the threshold of $0.25$ found with the specific setup of \secref{subsec:theory_axi}). We observed (not shown) that RK modes are present in all of the simulations that meet both this criterion and $Fr\geq 1$. This condition is not, however, sufficient for superrotation. Examination of the EMFC cospectra in subrotating simulations that meet both criteria show that they do feature RK modes, but there are generally too weak in face of the westward acceleration provided by the mean flow and other transient eddies. In the low-$T_\mathrm{rad}$ regime, this likely happens as the strong thermal damping significantly weakens the RK modes.

\cite{Wang2014} discussed the existence of RK instability in terms of two parameters: the Froude number Fr, and a midlatitude Rossby number $Ro(\phi_0)$ ($= Ro_T U(\phi_0)$ with our nondimensional scales). They found that the growth rate of RK modes was an increasing function of the latter, with positive growth rates for $Ro(\phi_0) \geq 0.1$, although they did not provide a physical explanation for this criterion. We nevertheless show the values of $Ro(\phi_0)$ in Fig. \ref{fig:Fr_Ro}c, as it appears that $Ro(\phi_0) \gtrsim 0.12$ is the best-fitting criterion for the existence of superrotation in our simulations. Further study is needed to understand why $Ro(\phi_0)$ is a better control on superrotation than the physical overlap of Rossby and Kelvin waves, as measured by $L_d/\phi_0$.

To summarize, RK modes are a necessary (but not always sufficient) condition for the appearance of superrotation in the configurations we have examined. Criteria based on the phase speed ratios and the spatial overlap between midlatitude Rossby waves and equatorial Kelvin waves help characterize in which parts of the parameter range these modes are present.

\section{Transition from a Slow Rotator to a Tidally-Locked Planet} \label{sec:transition}

We have seen that the mechanisms of superrotation on tidally-locked and axisymmetrically-forced planets bear some similarities. They both involve a coupling between midlatitude Rossby waves and equatorial Kelvin waves, but the waves in the tidally-locked case are predominantly forced stationary waves whereas those on axisymmetrically-forced planets arise from an instability and have a fast eastward propagation. The latter waves are also involved in the spinup of superrotation in some tidally-locked cases. In this section, we explore the transition between the axisymmetrically-forced and tidally-locked regimes. Specifically, we fix all parameters and run a suite of simulations where we only vary the longitudinal structure of the thermal forcing, forming a continuum between axisymmetric and tidally-locked thermal forcing. It is unlikely that any planet could in reality undergo such a transition: the switch to tidal locking would likely be accompanied with a change in the rotation rate, hence in $Ro_T$. Rather, we aim to understand the interplay between RK waves and stationary waves forced by zonal asymmetries.

All six simulations presented in this section feature $\mathcal{S}=0.05$, $E = 0.02$, $Ro_T=10$, and $T_\mathrm{rad} = 250$. We choose these values of $Ro_T$ and $T_\mathrm{rad}$ because both tidally-locked and axisymmetrically-forced cases superrotate in this regime, and the tidally-locked case also features RK instability during its spinup phase. 
The runs only differ in the zonal structure of their thermal forcing. The two end-members were already described in section \ref{sec:nonlinear}: one is axisymmetric (i.e., $\theta_{iE} = (1-\mathcal{S}\ln\Pi_i)\,\cos\phi\,(1/\pi)$) and the other one is tidally-locked (i.e., $\theta_{iE} = (1-\mathcal{S}\ln\Pi_i)\,\cos\phi\,\max(0,\cos\lambda)$). Four runs with increasingly stronger zonal asymmetries bridge these two. Recalling that $\max(0,\cos\lambda)$ expands in Fourier series as $1/\pi + (\cos \lambda)/2 + \hdots$, we define a parameter $\epsilon$ such that $\theta_{iE}$ is given, in these runs, by
\begin{equation}
    \theta_{iE} = (1-\mathcal{S}\ln\Pi_i)\,\cos\phi\,
    \left(\dfrac{1}{\pi} + \epsilon \cos \lambda\right),
\end{equation}
and we use four values of $\epsilon$: $0.05,0.1,0.25$ and $0.5$. All simulations are run for 1000 rotation periods, with statistical equilibrium reached within 300.

\begin{figure}[!t]
\begin{center}
\includegraphics[width=0.6\textwidth]{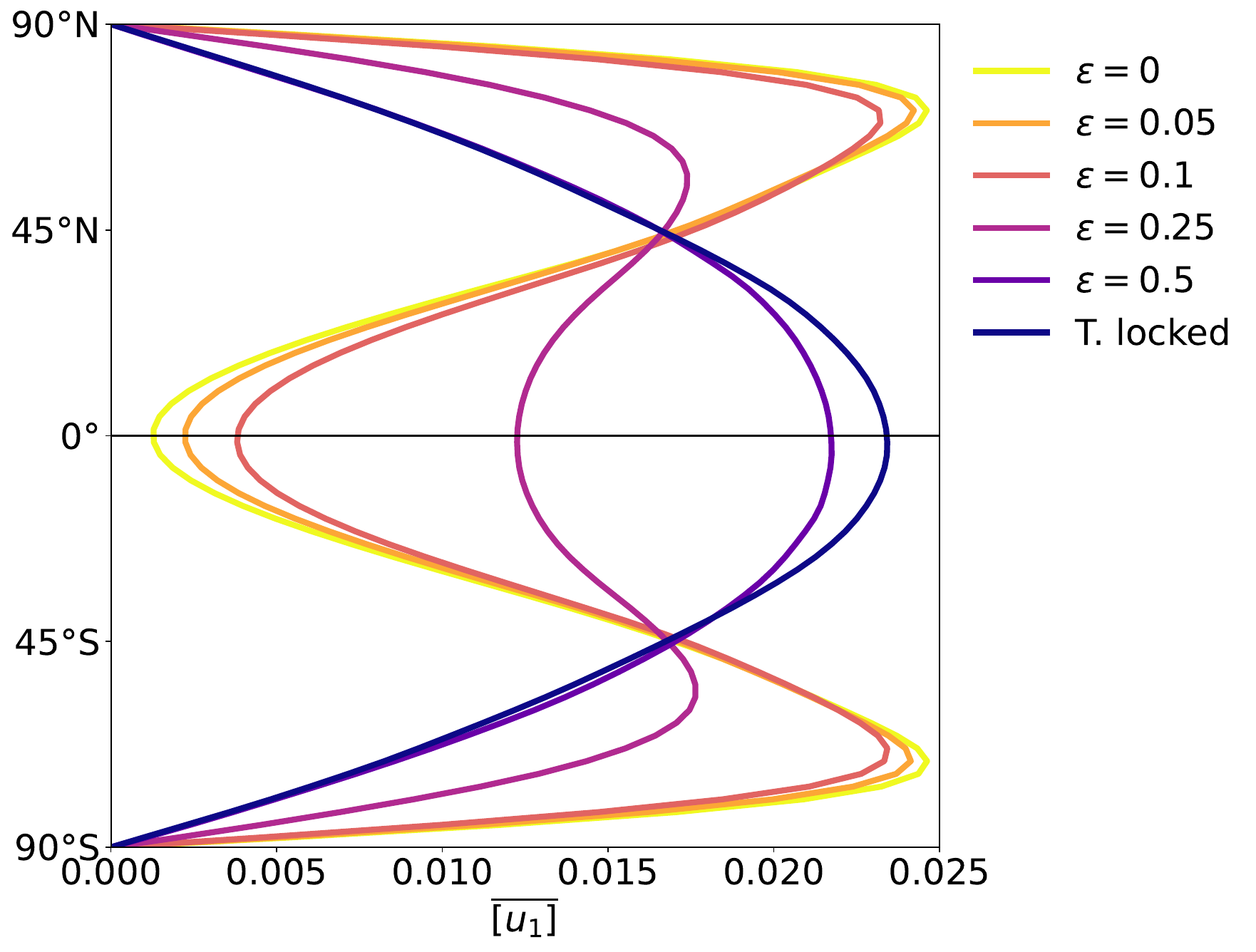}
 \caption{Upper layer zonal-mean zonal wind in a suite of two-level runs with increasing $n=1$ heterogeneities in thermal forcing, with $Ro_T= 10$ and $T_\mathrm{rad} = 250$.}\label{fig:utransition}
\end{center}
\end{figure}
Equilibrated $\overline{u_1}$ profiles (Figure \ref{fig:utransition}) observe a transition from a regime with two strong high-latitude jets and weaker equatorial flow (axisymmetric or weakly zonally asymmetric cases) to a regime where the zonal wind increases monotonously from the poles to the equator ($\epsilon = 0.5$ and tidally-locked cases). The jet strength increases with $\epsilon$, showing that stationary eddy forcing helps achieve stronger superrotation than RK instability alone. Finally, the $\epsilon = 0.5$ and tidally-locked simulations have a very similar profile, indicating that the higher-order harmonics in the zonal structure of the thermal forcing in this tidally-locked case only play a weak role. This also justifies our approach of studying the transition, by only varying the strength of the $n=1$ component of the forcing.

\begin{figure*}[t]
\begin{center}
 \includegraphics[width=\textwidth]{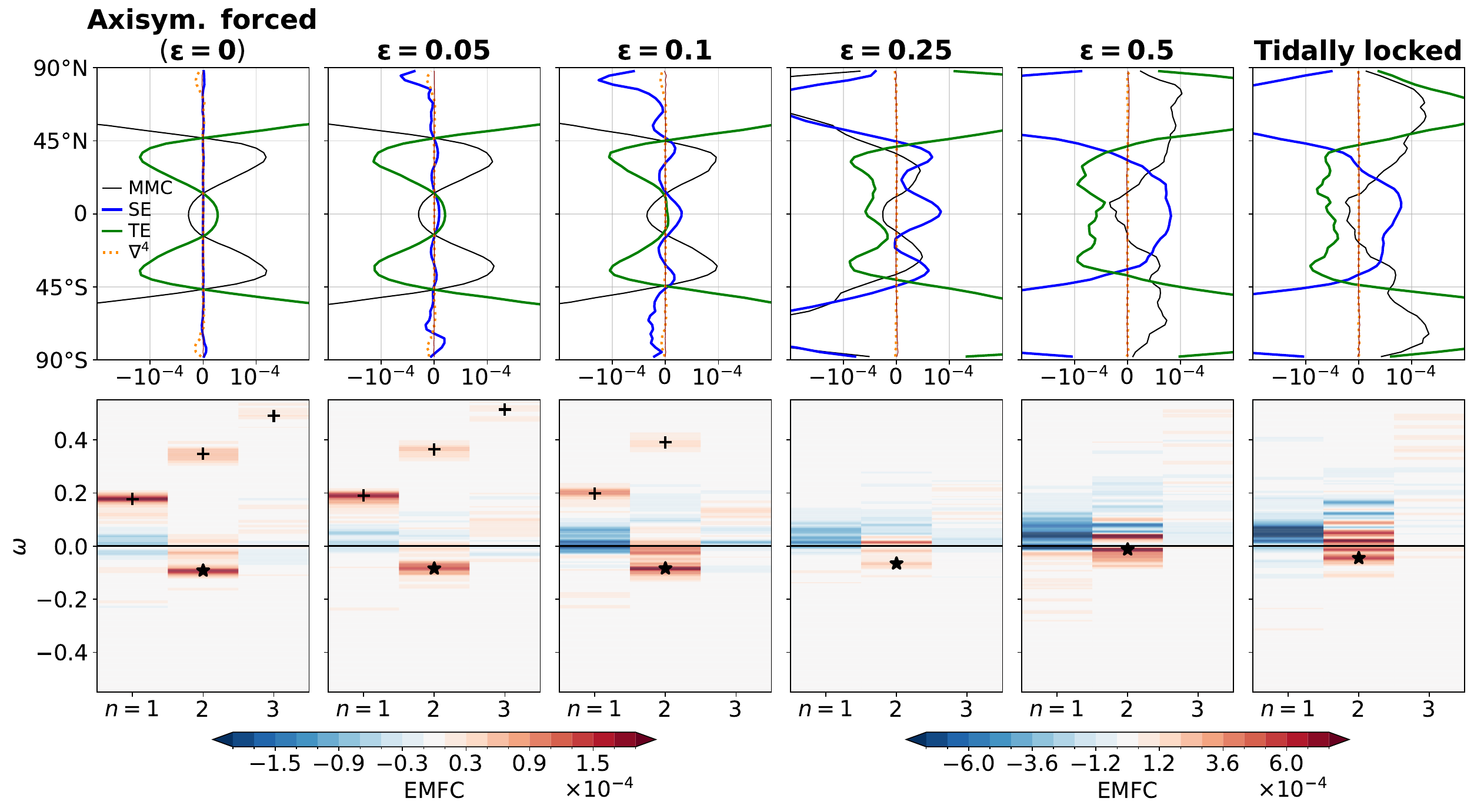}
 \caption{Contributions to the zonal-mean zonal momentum balance in the suite of two-level runs with increasing $n=1$ heterogeneities in thermal forcing. \textbf{(Top row)} Zonal-mean zonal momentum balance \eqref{eqn:mmc_se_te}, as in Fig \ref{fig:lowRoT}b,e. \textbf{(Bottom row)} Spectral decomposition of the EMFC due to transients $K_{n,\omega}(\phi)$, averaged for $\phi\in$ [10°S, 10°N]. Plus signs indicate RK modes, and stars indicate westward-propagating MRG-like waves. Gaussian smoothing with standard deviation $5\times 10^{-3}$ is applied along the frequency axis. Note that the color scale of the left three panels differs from that of the right three panels by a factor 4.}\label{fig:emfc_transition}
\end{center}
\end{figure*}

\Figref{fig:emfc_transition} (top row) shows the zonal-mean zonal momentum budget terms in statistical equilibrium. In all simulations, the mean meridional circulation accelerates $\overline{u_1}$ westward at the equator, due to the ascending branch of the Hadley circulation advecting weaker angular momentum from the lower layer. 

In the $\epsilon=0$ case, the absence of forced zonal heterogeneity mandates that the stationary eddy component vanish. Hence, the only term that can balance westward acceleration of the mean flow is that arising from transient eddies. The bottom row of \figref{fig:emfc_transition} explores the transient eddy component by way of the spectral decomposition \eqref{eqn:Knomega}. This quantifies the contribution of each wave of zonal wavenumber $n$ and frequency $\omega$ to transient EMFC at each latitude, here averaged over the equatorial band. Several isolated peaks of eastward acceleration stand out, the strongest corresponding to $n=1$ and $\omega = 0.2$: this is the $n=1$ RK mode. Higher-order RK modes are also present and are flagged with plus signs on the spectral diagram. Another strong peak, marked with a star, corresponds to a $n=2$ westward-traveling wave; further investigation shows that its structure resembles a MRG wave.

As $\epsilon$ increases, introducing stationary zonal inhomogeneity, the westward equatorial acceleration resulting from the mean meridional circulation changes little. However, the stationary component progressively replaces transient eddies in providing eastward acceleration, until the transient eddy component switches sign between $\epsilon=0.1$ and $\epsilon=0.25$. RK modes are present in statistical equilibrium for $\epsilon\leq0.1$. Beyond that, the equatorial flow strengthens significantly (Fig. \ref{fig:utransition}), so that the Froude number is too small for RK modes to persist. While this may suggest that stationary eddies are the sole driver of superrotation for $\epsilon\geq 0.25$, transient eddies are still found to play a prominent role. Indeed, RK modes are observed during the spinup phases of the $\epsilon\geq 0.25$ and tidally-locked simulations (shown for the latter in \figref{fig:RK_tidally_locked}). MRG waves also persist in these simulations (Fig. \ref{fig:emfc_transition}) and have a strong contribution towards eastward equatorial acceleration. Westward acceleration is primarily the result of low-frequency wavenumber 1 eddies, which are present in all simulations and strengthen with $\epsilon$. 

The simulations of \cite{Suarez1992} and \cite{Saravanan1993} bear some comparison to those here. Using a similar two-level model but with Earth-like parameters (i.e. $Ro_T\simeq 0.1$, so relatively quickly rotating), they studied the effect of increasing the strength of stationary $m=2$ thermal forcing at the equator. They observed a sudden switch from subrotation to superrotation that seemed primarily driven by the behavior of transient eddies. \cite{Kraucunas2005} argued that some of this behavior was not robust, as similar experiments in more vertically-resolved GCMs showed stationary eddies to have the prominent role. 
Our simulations behave rather similarly to multi-level GCMs, as stationary eddy forcing (rather than transient eddy forcing) increases with increasing zonally asymmetric heating. This may also explain why the transition in equatorial winds is not as sudden in our simulations (Fig. \ref{fig:utransition}) as those of \cite{Suarez1992} and \cite{Saravanan1993}: their mechanism relied on a change in the low-latitude absorption of eddies generated at high latitudes, which has a threshold behavior depending on the equatorial wind speed. While the behavior of our 2-level runs is reassuring in that aspect, it does not obviate the necessity of verifying the results in multilevel models.

\section{Discussion and Conclusions} 
\label{sec:conclusions}

In this paper we have explored the presence and mechanisms of superrotation on both tidally-locked planets and axisymmetrically-forced slow rotators, using perhaps one of the simplest models fit for that purpose: a two-level primitive equation model. Four parameters control its behavior: a thermal Rossby number $Ro_T$, which is higher for small, slowly-rotating, or strongly irradiated planets; a nondimensional thermal relaxation time scale $T_\mathrm{rad}$, which is a proxy for optical thickness; an Ekman number $E$ parameterizing surface drag, and a nondimensional thermal stratification $\mathcal{S}$. The emphasis is placed on the behavior of superrotation as a function of the first two, although there is some dependence on $E$. 

The steady linear response to the day-night insolation contrast on tidally-locked planets, similar to that of the Matsuno--Gill model, provides the basic organizing structure for the eddy effects. However, a Matsuno--Gill pattern of itself, in a single layer model, is insufficient to produce superrotation, and both surface drag and baroclinicity (to allow for vertical fluxes of momentum) are needed. In the presence of surface drag the strength of the eddy-momentum flux convergence (EMFC) is a decreasing function (both in absolute terms and relative to midlatude EMFC peaks) of both $Ro_T$ and $T_\mathrm{rad}$. That is, slower rotation and a thicker atmosphere both inhibit superrotation, insofar as the quasi-liner results are relevant. 

In slowly-rotating axisymmetrically-forced planets, unstable Rossby-Kelvin (RK) modes, which qualitatively resemble the Matsuno--Gill pattern, similar to those previously identified in GCMs and shallow water models, also produce superrotation, again provided some vertical structure is present.

Fully nonlinear integrations reveal very rich behavior. Consistent with the quasi-linear results from the Matsuno--Gill structure, superrotation in tidally-locked planets is less favored as $Ro_T$ and $T_\mathrm{rad}$ increase when $Ro_T\leq 1$. However, at higher levels of nonlinearity the dependence on $Ro_T$ switches. The RK modes, generally thought to be more relevant in the context of axisymmetrically-forced planets, arise during the  the spinup of superrotation, even in tidally-locked planets. A broad class of tidally-locked planets with high $T_\mathrm{rad}$ (i.e., a thicker atmosphere) subrotate; the propensity to subrotate also increases when surface drag is removed, although only for $Ro_T\leq 1$. Thus, although the structure of the eddies is set by the directly forced Matsuno--Gill pattern, transients play a major role in setting the equilibrated state of most tidally-locked runs. 

On axisymmetrically-forced planets, superrotation mainly appears for $Ro_T\geq 1$, i.e., for slow rotators. The factor that limits its appearance is the spatial overlap between equatorial Kelvin waves and midlatitude Rossby waves. If the midlatitude jets are too far poleward or the Kelvin waves too equatorially confined, the overlap is too weak to produce superrotation. Also, superrotation is inhibited if the midlatitude jets produce Rossby waves that propagate and break in equatorial regions, a phenomenon that seems likely to occur on more Earth-like planets when $Ro_T$ is small.

We finally presented a continuum of simulations bridging axisymmetrically-forced and tidally-locked states by applying progressively stronger wavenumber-1 thermal forcing, in a regime where RK modes (and superrotation) are present in the two end-members.  In spite of a seemingly continuous transition in wind patterns when transitioning from an axisymmetrically-forced  to a tidally-locked planet, the processes responsible for superrotation switch quickly when weak zonal asymmetries are applied. Specifically, a stationary Gill-like pattern soon dwarfs the contribution from propagating waves in the axi-symmetric case in setting the total eddy momentum fluxes. However, the two processes can and do coexist, and some of the unstable modes from the axi-symmetic case may still be present when strong zonal asymmetries are applied. RK modes are present in the spinup phases of all cases, but are only present in the equilibrated state for weak zonal inhomogeneity in the forcing.

Overall, it may be fairly said that the mechanisms of superrotation in slowly-rotating planets are becoming well-established, with a Rossby--Kelvin instability involving both horizontal and vertical eddy momentum fluxes playing a role. The (admittedly limited) observations of slow rotators in the Solar System are generally supportive of the mechanisms identified in this paper and previous investigators \citep[e.g.,] []{Iga2005, Mitchell2010, Wang2014, ZuritaGotor2018}.  Tidally-locked planets present a different challenge, because although there are many examples, detailed observations are sparse and will remain so, even with JWST. Our results suggest that superrotation on such planets is ubiquitous but not universal.  The importance of drag at the base of the moving atmosphere suggests that the interaction of the shallow atmosphere (i.e., that driven by stellar irradiation) with the deeper atmosphere deserves further investigation, for the mechanisms producing such a drag are not obvious. 

Finally, to comment on our model, the use of just two levels is motivated by the requirements of having (ideally) both completeness of mechanisms and conceptual simplicity, enabling  tractability in analysis as well as realistic behavior. However, that is not always possible and the strong vertical truncation of the model may misrepresent vertical momentum transport by the eddies and the mean flow alike. Future work will explore whether the mechanisms and phenomena seen here hold when the vertical structure is much better resolved.

\begin{acknowledgments}
The authors would like to thank Keaton Burns and Daniel Lecoanet for their invaluable help with Dedalus, and Jonathan Mitchell and Tad Komacek for a number of useful conversations about superrotation. Both authors were supported by the 2023 WHOI Geophysical Fluid Dynamics Summer Program (funded by the National Science Foundation and the Office of Naval Research), where this project was started. QN was partially supported by an ETH Z{\"u}rich Postdoctoral fellowship (Project No. 24-1 FEL-032). For the purpose of open access the authors have applied a Creative Commons Attribution (CC BY) license to any Author Accepted Manuscript version arising from this submission.

\end{acknowledgments}

\begin{contribution}

Conceptualization: G. K. Vallis \& Q. Nicolas
Investigation: Q. Nicolas
Writing – original draft: Q. Nicolas
Writing – review \& editing: G. K. Vallis \& Q. Nicolas.


\end{contribution}

%

\software{Code used to run the Dedalus simulations, processed simulation output, and code used in producing the figures will be archived at Zenodo upon completion of the review process.}


\appendix

\section{Drag-free solution of the two-level Gill problem} 
We seek a solution of the linear system \eqref{eqn:mom_lin_1}--\eqref{eqn:thermo_lin_2} in the case where $E=0$. The vorticity balance in layer $i$ is
\begin{equation}\label{eqn:AppA-vorticity}
\hat f \nabla\cdot \ubb_i + \beta v_i = 0,
\end{equation}
where $\beta = \partial_\phi \hat f = \cos \phi$. Combining with continuity, one obtains two relationships. The first one is $v_1+v_2 = 0$, which also implies $u_1+u_2 = 0$ by continuity; hence $\ubb_1 = -\ubb_2$. The second relationship is
\begin{equation}\label{eqn:AppA-1}
\beta(v_1 - v_2) = 4\hat f \omega .
\end{equation}
Combining the zonal momentum equations of both layers and the hydrostatic equation, one also obtains
\begin{equation}\label{eqn:AppA-2}
\hat f (v_1 - v_2) = \dfrac{1}{\cos \phi} \partial_\lambda (\Phi_1 - \Phi_2) = \dfrac{\gamma}{\cos \phi} \partial_\lambda (\theta_1 + \theta_2).
\end{equation}
Finally, summing the thermodynamic equations of both layers gives 
\begin{equation}\label{eqn:AppA-4}
-2\mathcal{S}Ro_T\omega = \dfrac{(\theta_{1E} + \theta_{2E}) - (\theta_1 + \theta_2)}{T_\mathrm{rad}}.
\end{equation}
Combining \eqref{eqn:AppA-1}-\eqref{eqn:AppA-4} and assuming that each scalar field $\varphi$ has a structure $\varphi(\phi,\lambda) = \tilde \varphi(\phi) e^{in\lambda}$ ($n=1$ for the problem considered in section \ref{subsec:theory_tl}), a closed solution for $\theta_1 + \theta_2$ emerges:
\begin{equation}
\tilde\theta_1 + \tilde\theta_2 = \dfrac{\tilde\theta_{1E} +\tilde \theta_{2E}}{1 - \frac{in\gamma\mathcal{S}Ro_T T_\mathrm{rad}}{2\hat f^2}}.
\end{equation}
The meridional wind fields can in turn be obtained using \eqref{eqn:AppA-2}, and the zonal wind from the meridional momentum equations. Noting that $u_1+u_2=0$, upper level EMFC \eqref{eqn:EMFC1} is
\begin{equation}
    \mathrm{EMFC}_1 = - Ro_T \left( \pp {\overline{u_1v_1}} \phi - 2 \tan\phi \overline{u_1v_1}\right).
\end{equation}
Noting that $\overline{u_1v_1} = \frac{1}{2}\Re(\tilde u_1 \tilde v_1^*)$, a little algebra yields
\begin{equation}\label{eqn:nodrag_EMFC}
\mathrm{EMFC}_1 = - Ro_T \kappa \dfrac{ n \gamma^2 \xi^2 }{8} \dfrac{Y (Y^3 - 5 Y^2 - 7 \kappa^2 Y + 3  \kappa^2)\sqrt{1-Y}}{(\kappa^2 + Y^2)^3},
\end{equation}
where $Y = \sin^2\phi$, $\kappa = n \gamma \mathcal{S} Ro_T T_\mathrm{rad} / 2$ and $\xi = 1 - S(\ln(\Pi_1)+\ln(\Pi_2)/2 \simeq 1$.

How does the magnitude of EMFC$_1$ depend on the input parameters $\mathcal{S}$, $Ro_T$, and $T_\mathrm{rad}$? When $\kappa \gg 1$, one sees from \eqref{eqn:nodrag_EMFC} that $|$EMFC$_1|$ $\propto Ro_T \kappa \kappa^2/\kappa^6 \propto Ro_T^{-2} \mathcal{S}^{-3} T_\mathrm{rad}^{-3}$. When $\kappa \ll 1$, the maximum of \eqref{eqn:nodrag_EMFC} is attained for $Y\simeq \kappa$, and $|$EMFC$_1|$ $\propto Ro_T \kappa {\kappa^3}/{\kappa^6} = Ro_T \kappa^{-2} \propto Ro_T^{-1} \mathcal{S}^{-2} T_\mathrm{rad}^{-2}$. With $n=1$ and $\gamma\simeq 0.12$, the transition between these scalings happens for $\mathcal{S}Ro_T T_\mathrm{rad} \sim 20$.

Last, we provide an expression for the low-level zonal wind $u_2$ at the equator:
\begin{equation}
u_2 (\phi=0,\lambda) = -\dfrac{2\xi}{n \mathcal{S} Ro_T T_\mathrm{rad}}\sin(n\lambda)
\end{equation}

\section{Eddy momentum flux convergence in the general two-level Gill problem} 
We move away from the simpler drag-free case, and seek to obtain a simplified expression for the upper-level EMFC at the equator in the 2-level Gill model. As explained in section \ref{sec:theory}, EMFC$_1 = -Ro_T \overline{\omega (u_2-u_1)} = -Ro_T\overline{\omega u_2}$, because $\overline{\omega u_1} = 0$ due to the absence of drag in the upper layer. From \eqref{eqn:AppA-4} (which is valid whatever the value of $E$), one may derive  
\begin{equation}
    -Ro_T\overline{\omega u_2} = \dfrac{1}{\mathcal{S} T_\mathrm{rad}}\overline{(\theta_{1E}+\theta_{2E})u_2} - \dfrac{1}{\mathcal{S} T_\mathrm{rad}}\overline{(\theta_{1}+\theta_{2})u_2}.
\end{equation}
We proceed to show that $\overline{(\theta_{1}+\theta_{2})u_2} = 0$.
Subtracting the upper-layer zonal momentum balance from the lower-layer one at the equator (where $v_1 = v_2 = 0$), one obtains 
\begin{equation}
\partial_\lambda (\Phi_2-\Phi_1) + E u_2 = 0.
\end{equation}
Combining with hydrostasy, 
\begin{equation}
-\gamma \partial_\lambda(\theta_{1}+\theta_{2} )+ E u_2 = 0.
\end{equation}
Hence, 
\begin{equation}
\overline{(\theta_{1}+\theta_{2})u_2} \propto \overline{(\theta_{1}+\theta_{2})\partial_\lambda(\theta_{1}+\theta_{2} )} = 0
\end{equation}

\section{Estimation of planetary parameters}
The planets shown in Figure \ref{fig:regime_diagram_tl} are selected following \cite{Pierrehumbert2019} and \cite{Perezbecker2013}. $\Delta\Theta_h$ is taken as the equilibrium temperature $T_\mathrm{eq}$ of the planet. $c_p$ is $10^3$ J kg$^{-1}$ K$^{-1}$ for terrestrial planets \citep[assuming a N$_2$ atmosphere, which is in no way certain, e.g.,][]{Hammond2017} and $1.2\times10^4$ J kg$^{-1}$ K$^{-1}$ for hot Jupiters (i.e., assuming a H$_2$ atmosphere). Following \cite{Perezbecker2013}, the radiative relaxation time scale is estimated as
\begin{equation}
    \tau_\mathrm{rad} = \dfrac{Pc_p}{4g\sigma T_\mathrm{eq}^3},
\end{equation}
where $\sigma$ is is the Stefan–Boltzmann constant and $g$ is the surface gravity. $P$ is the atmospheric depth on tidally locked planets, and can be taken at a representative emission level or somewhat deeper for hot Jupiters. We take it as 1 bar in both cases, although there is considerable uncertainty about this figure. Data for terrestrial planets are from \cite{Pierrehumbert2019}, \cite{Xue2024}, \cite{Cadieux2024}, \cite{Agol2021}, \cite{Bonomo2025} and \cite{Bourrier2018}. All data for hot Jupiters are from \cite{Stassun2017}, except for $T_\mathrm{eq}$ which is from \cite{Perezbecker2013}.

For Fig. \ref{fig:regime_diagram_axi}, data for Earth are standard, and $\tau_\mathrm{rad}$ is taken as 40 days \citep[e.g.,][]{Held1994}. For Mars, we use $c_p = 736$ J kg$^{-1}$ K$^{-1}$, $\tau_\mathrm{rad}$ = 2 days, and $\Delta \Theta_h = 300 $ K \citep{Haberle1997}. For Titan, we use $c_p = 10^3$ J kg$^{-1}$ K$^{-1}$ and $\Delta \Theta_h = 20 $ K \citep{Mitchell2016}. $\tau_\mathrm{rad}$ is taken as $3\times 10^8$ s \citep{Bezard2018}.

All values are summarized in Table \ref{tab:planetary_parameters}.

\begin{table}[t]
{\footnotesize
\begin{tabular}{@{}lcc}
Planet & $Ro_T$ & $T_\mathrm{rad}$ \\
\hline
GJ1132b      & 1.7 & 5.6\\
LHS1140 b    & 76 & 4.5 \\
Trappist 1b  & 1.2 & 23 \\
Trappist 1c  & 2.7 & 23 \\
Trappist 1d  & 12 & 37 \\
55 Cancri e  & 0.48 & 0.20 \\
Kepler 10b   & 1.2 & 0.19 \\
\end{tabular}
\begin{tabular}{lcc}
Planet & $Ro_T$ & $T_\mathrm{rad}$ \\
\hline
HD-189733b & 0.51 & 9.2 \\
HD-209458b & 1.0 & 7.7 \\
HD-149026b & 2.7 & 3.6 \\
HAT-P-7b   & 0.52 & 1.6 \\
WASP-18b   & 0.16 & 0.29 \\
WASP-12b   & 0.10 & 4.1 \\
&&\\
\end{tabular}
\begin{tabular}{lcc}
Planetary body & $Ro_T$ & $T_\mathrm{rad}$ \\
\hline
Earth      & 0.07 & 500 \\
Mars       & 0.32 & 24 \\
Titan      & 36 & 2900 \\
&&\\
&&\\
&&\\
&&\\
\end{tabular}
}
\caption{Thermal Rossby number and nondimensional radiative time constant for various tidally-locked exoplanets and Solar System planetary bodies.}
\label{tab:planetary_parameters}
\end{table}

\section{Spectral Decomposition of the Transient Eddy Momentum Flux Convergence}
We show how the TE term in \eqref{eqn:mmc_se_te} is decomposed into contributions from eddies of different properties. The transient part of a given field $A$ is Fourier-transformed in longitude and time: 
\begin{equation}
    A^\dagger(t,\lambda,\phi) =  \Re\displaystyle\sum_{n=0}^N e^{in \lambda}\int_{-\infty}^{+\infty} \tilde A(\omega,n,\phi) e^{-i\omega t} \mathrm{d}\omega,
\end{equation}
where $N$ is the maximum resolved zonal wavenumber. The correlation between two fields $A^\dagger$ and $B^\dagger$ is decomposed as 
\begin{equation}
    \overline{[A^\dagger B^\dagger]} = \dfrac{1}{2} \displaystyle\sum_{n=0}^N \int_{-\infty}^{+\infty} \Re(\tilde A(\omega,n,\phi) \tilde B^*(\omega,n,\phi)) \mathrm{d}\omega,
\end{equation}
This way, we decompose the transient EMFC at each latitude as 
\begin{equation}\label{eqn:Knomega}
\begin{aligned}
\text{TE}(\phi) &=  Ro_T\left(\overline{\left[\zeta_1^\dagger v_1^\dagger\right]} - \overline{\left[\omega^\dagger (u_2-u_1)^\dagger\right]}\right)\\
&= \displaystyle\sum_{n=1}^N \int_{-\infty}^{+\infty} K_{n,\omega} (\phi) \mathrm{d}\omega,
\end{aligned}
\end{equation}
where 
\begin{equation}
K_{n,\omega} (\phi) = \dfrac{Ro_T}{2}\Re\left(\tilde\zeta_{1}\tilde v_{1}^* + \tilde\omega(\tilde u_{1}^* - \tilde u_{2}^*)  \right)
\end{equation}
denotes the contribution of waves of frequency $\omega$ and zonal wavenumber $n$ to EMFC at latitude $\phi$. The contribution from wavenumber 0 (i.e., time variations in the zonal mean flow) has been dropped, as it is negligible in these simulations. In Figure \ref{fig:emfc_transition}, we show averages of $K_{n,\omega} (\phi)$ over the equatorial region.

In \secref{subsec:2levTL}, we further decompose $TE(\phi)$ as
\begin{equation}\label{eqn:lat_c_decomp}
\text{TE}(\phi) = \displaystyle \int_{-\infty}^{+\infty} \sum_{n=1}^NK_{n,c} (\phi) \mathrm{d}c,
\end{equation}
where $c = \omega \cos\phi / (n Ro_T)$ is the phase speed (the factor $1/Ro_T$ comes from our choice of nondimensionalization for velocities), and $K_{n,c} = (nRo_T/\cos\phi)K_{n,\omega}$. Figures \ref{fig:lowRoT}c,f and \ref{fig:highRoT}c,f show $\sum_{n=1}^NK_{n,c} (\phi)$ as a function of $c$ and $\phi$.

\newpage
\bibliographystyle{apalike}
\bibliography{references}  



\end{document}